\documentstyle[12pt,aasms4]{article}
\def\clock{\count0=\time \divide\count0 by 60
     \count1=\count0 \multiply\count1 by -60 \advance\count1 by \time
     \number\count0:\ifnum\count1<10{0\number\count1}\else\number\count1\fi}

\begin{document}
\title{A High-resolution  Adaptive Moving Mesh Hydrodynamic Algorithm}
\author{Ue-Li Pen\altaffilmark{1}}
\affil{Harvard Society of Fellows and Harvard-Smithonian Center 
for Astrophysics\\
Cambridge, MA 02138 }
\altaffiltext{1}{e-mail I: upen@cfa.harvard.edu}
\newcommand{\rg}{\sqrt{g}}
\newcommand{\etal}{{\it et al }}
\newcommand{\Nabla}{\bigtriangledown}

\begin{abstract}

An algorithm for simulating self-gravitating cosmological
astrophysical fluids is presented.  The advantages include a large
dynamic range, parallelizability, high resolution per grid element and
fast execution speed.  The code is based on a finite volume flux
conservative Total-Variation-Diminishing (TVD) scheme for the shock
capturing hydro, and an iterative multigrid solver for the gravity.
The grid is a time dependent field, whose motion is described by a
generalized potential flow.  Approximately constant mass per cell can
be obtained, providing all the advantages of a Lagrangian scheme.  The
grid deformation combined with appropriate limiting and smoothing
schemes guarantees a regular and well behaved grid geometry, where
nearest neighbor relationships remain constant.  The full hydrodynamic
fluid equations are implemented in the curvilinear moving grid,
allowing for arbitrary fluid flow relative to the grid geometry.  This
combination retains all the advantages of the grid based schemes
including high speed per fluid element and a rapid gravity solver.

The current implementation is described, and empirical simulation
results are presented.  Accurate execution speed calculations are
given in terms of floating point operations per time step per grid
cell.  This code is freely available to the community.

\end{abstract}

\section{Introduction}

Astrophysical hydrodynamics is characterized by a large range in
density, temperature and length scales, where strong shocks often play
an important role.  This poses a great challenge to attempts at
simulating such processes numerically.  Traditionally, simulations
have either been carried out on a static mesh (Cen \etal 1990, Cen
1992), or using Monte-Carlo techniques by following particle
trajectories in Smooth Particle Hydrodynamic (SPH) models (Evrard
1988, Hernquist and Katz 1989, Navarro and White 1993, Monaghan 1995).

There are two main difficulties in simulating astrophysical fluids
numerically.  The first is the fact that often the flows occur at very
high mach number, leading to frequent development of strong shock
discontinuities.  The second difficulty is the large range in length
scales involved when gravitational clustering occurs.  The mesh
schemes can often address the first problem very well through the use
of the Total Variation Diminishing (TVD) paradigm, while the particle
methods have been primarily developed to address the second
problem. In this paper we will describe a code which attempts to
address both problems.

The advantages of mesh-based TVD approaches (Yee 1989) include the
implementation of modern hydrodynamic concepts based on the
characteristic field decomposition.  The general family includes the
Piecewise Parabolic Mesh (PPM) (Collella and Woodward 1984) and Harten
schemes (Harten 1983) which have been successfully applied to
cosmological hydrodynamics (Ryu \etal 1993, Bryan \etal 1994).  These
provide for high resolution capturing of shock fronts in 1--2 cells
and high order accuracy away from extrema.

A finite difference scheme is difficult to implement across
discontinuities, where the differential equation becomes ill-defined,
and requires a mathematical treatment in terms of internal boundaries.
In order to obtain a meaningful convergent result, the classical
treatments added large amounts of artificial viscosity and diffusion,
which prevents the formation of discontinuities on scales shorter than
a cell size.  At the same time, such a large viscosity severely
degrades the resolution of the simulation.  Modern shock capturing
approaches contain two ingredients.  They express the fluid equations
in integral flux conservative form.  This is accomplished by dividing
space into a set of control volumes, in the simplest case by a
Cartesian cubical lattice.  On the boundary between volumes, one
calculates the flux which passes between cells.  Whatever flux is
taken out of one cell is always added to its neighboring cell.  Using
this approach, one automatically satisfies the Rankine-Hugionot
conditions, and is thus guaranteed the correct shock jump conditions
and shock propagation speed.  The second ingredient is a flux or slope
limiter.  This replaces the traditional artificial viscosity.  By
analyzing the characteristics of the hyperbolic PDE, one obtains
constraints on the flux functions, which causes them to remain well
behaved in the presence of discontinuities.  This prevents
instabilities and post shock oscillations.  The characteristic
decomposition allows a high resolution capturing of discontinuities,
often in two or fewer cells.  An alternate view point is to describe a
flux limiter as a strongly nonlinear viscosity scheme, which adds just
enough diffusion to prevent numerical instabilities.

The simplest way to implement these flux conservative high resolution
TVD schemes is in fixed regular Eulerian coordinates which are uniform
in space (Ryu et al 1993).  Gravitational instability drives fluids to
collapse to very dense configurations.  The cores of clusters of
galaxies are overdense by $10^3-10^4$ of the mean density of the
universe.  Often one is interested in the physical processes occurring
in these dense regions, for example the X-ray properties.  In a fixed
Eulerian mesh a large fraction of the mass ends up in a small fraction
of the grid cells, leading to a degradation in resolution.  The
advantage of such Eulerian approaches include simplicity of
implementation, high computational speed per grid cell,
straightforward data parallel implementation on distributed memory
computers, and high resolution of shocks.  Simulations are usually
limited by the amount of available memory.

In many problems of astrophysical interest, the  physical processes
occur on constant mass scales, which argues in favor of constant mass
resolution algorithms.  The simplest way of building a constant mass
resolution scheme is by utilizing a Lagrangian coordinate system, where
the numerical control points are frozen into the fluid.  An additional
problem arises in such an approach.  Any rotation in the fluid tends to
move cells which are initially close to each other to large
separations, causing a rearrangement of nearest neighbor relationships.

One popular approach, SPH, addresses this problem by resampling the
list of nearest neighbors at each time step.  Unfortunately, it is
then no longer possible to maintain the integral flux-conservative
control volume and characteristic TVD approach.  In order to conserve
mass, SPH further interprets each numerical grid point to be a fuzzy
particle of constant mass.  The density is then defined as a
statistical quantity which is estimated by calculating the distance to
the nearest 30--100 neighbors.  Artificial viscosity is used to
prevent the formation of discontinuities.  In each case, there is a
trade off between accuracy and resolution.  By smoothing over more
neighbors, one obtains a more accurate estimate of the density field,
which is limited by the $\sqrt{N}$ Poisson noise, while also reducing
resolution due to the same smoothing.  A similar effect holds for
viscosity.  An additional problem arises when the list of nearest
neighbors is determined using a spherical search algorithm.  The
nearest neighbor distribution could be highly anisotropic, which further
degrades the resolution.  This issue has been addressed by Martel and
Shapiro in ASPH (Shapiro et al 1995).  We conclude that SPH is a
Monte-Carlo approach, which is constrained by Poisson noise,
viscosity, anisotropy, and the cost of searching for nearest
neighbors.  Its primary advantages include mass based resolution which
allows a high range in spatial resolution, and
ease of implementation in a very large range of problems, including
problems with vacuum regions and complicated equations of state.  Its
cost is high computational effort and low resolution per particle.
Such simulations are usually CPU time limited on current
computers.

An alternate Lagrangian mesh approach has been developed by Gnedin (1995).
It forces the nearest neighbors to remain fixed in time.  If the grid
becomes excessively distorted, it reverts to an Eulerian scheme in the
fixed coordinate system.

Significant work on combining the advantages of these approaches has
been implemented by several authors.  Berger and Collela (1989)
developed a technique for local mesh refinement on regular meshes.
Lohner (1985) and Xu (1996) developed an unstructured grid, which
dynamically adds and removes nodes as necessary.  R. Fiedler and
Mouschovias (1992 hereafter FM; 1993) developed a moving mesh approach
for cylindrically symmetric Magnetohydrodynamics (MHD).  B. Fiedler
and Trapp (1992 hereafter FT, see also Trapp and Fiedler 1995) applied
a general curvilinear transform to model two-dimensional tornado
dynamics.  A review of many methods classified as
Node-Movement-Techniques is given by Thompson \etal\ (1985) and Hawken
(1991).

The purpose of this paper is to present a method called Moving Mesh
Hydro or MMH for short, with primary emphasis on speed and simplicity
for application to cosmological hydrodynamics.  It presents a general
framework which can be extended and improved in the future.  The
essential concept is to formulate a high resolution flux conservative
scheme on a general moving curvilinear coordinate system in the spirit
of FM and FT.  As in
their work, we attempt to follow only the divergence of the fluid flow
but not its vorticity.  Such a grid would resist any twisting or
shearing, and still maintain a constant mass per grid cell.  To
describe this mathematically, we recall that a general coordinate
transformation has three degrees of freedom.  By requiring
approximately constant mass per control cell, we have imposed one
constraint.  The remaining two degrees of freedom will be used to
prevent the appearance of vorticity in the grid motion.  Caustics and
discontinuities are treated by applying compression limiters and by
smoothing the deformation transformation.  This guarantees the
stability and regularity of the grid, and allows us to use the same
neighbors, volumes, and flux boundaries throughout the simulation.
The features thus include exact conservation form and resolution
proportional to density, which aims to combine the advantages of both
Eulerian and Lagrangian techniques.  (The cosmological energy equation
is not exact, but can be split into an exact hyrodynamic form with
graviational source term).  It features a low computational
cost and high resolution per grid cell.  The current implementation is
also quite memory efficient and parallelizable, which allows it to
make efficient use of present technology.

In the past, similar approaches became dominated by the cost of
computing the optimal grid configuration, which is a nonlinear global
optimization problem, and is often much more complex than the solution
of the hydrodynamic equations.  This paper presents the notion that the
astrophysically interesting constraint of maintaining constant mass per
cell can be achieved from an evolutionary point of view by solving only
linear elliptic equations at each time step.  The fast multigrid
solvers accomplish this in linear time, with effective speeds which are
competitive with fast Fourier transforms.  The cost over a fixed
Eulerian grid are thus the overhead for implementing the curvilinear
form of the equations as well as requiring two applications of the
Poisson solver.

\section{Formulation}

While the MMH algorithm is quite general and could be applied to any
three dimensional simulations (Pen 1995), we will examine the case of
systems of conservation laws, and in particular the Newtonian Euler
equation in detail.  We will consider only the case of cosmological
interest, where resolution will attempt to remain constant in mass.


Consider a numerical grid of coordinates
$\xi\equiv(\xi_1,\xi_2,\xi_3)$. In 
order to determine the physical position of each lattice point, one
needs to specify the Cartesian coordinate $x(\xi,t)$ of each
curvilinear coordinate.
We will borrow most of the notation from general relativity, which
provides a concise framework to describe general curvilinear coordinate
transformations.   We will only consider with three
spatial dimensions where the metric is positive definite, and the
underlying space is always Euclidean, i.e.  the Riemann tensor
vanishes everywhere.  Thus covariant derivatives always commute, and all
the nice properties of flat space hold.  The flat metric
$g_{ij}=\delta_{ij}$ is just the Kronecker delta function.  The
curvilinear metric is then
\begin{equation}
g_{\alpha\beta}=\frac{\partial x^i}{\partial \xi^\alpha}
            \frac{\partial x^j}{\partial \xi^\beta} \delta_{ij}.
\end{equation}
Repeated indices  obey the summation convention, which means that
they are dummy indices and should be summed from 1--3.  Latin indices
denote Cartesian coordinate labels $x^i$, while Greek indices imply
curvilinear coordinates $\xi^\alpha$.  A dot will imply partial
differentiation for time.  A comma will denote a partial derivative.

In Cartesian conservation form, the Euler equations for fluid dynamics are
\begin{eqnarray}
\frac{\partial\rho}{\partial t} + \frac{\partial}{\partial x^i}
\rho v^i &=&0 ,
\nonumber\\
\frac{\partial\rho v^i}{\partial t} + \frac{\partial}{\partial x^j}
\left[\rho v^i v^j + P \delta^{ij} + \frac{1}{4\pi G}\left(
\frac{\partial V}{\partial x^k} \frac{\partial V}{\partial
x^l}(\delta^{il} \delta^{jk}-\frac{1}{2}\delta^{ij}\delta_{kl}) \right)
\right]
 &=&0 ,
\nonumber\\
\frac{\partial e}{\partial t} + \frac{\partial}{\partial x^i}\left[
(e+\rho V+P) v^i + \frac{1}{8\pi G}(V \frac{\partial \dot{V}}{\partial x^i}
- \dot{V} \frac{\partial {V}}{\partial x^i}) \right] &=& 0 ,
\label{eqn:euler}
\end{eqnarray}
where the energy density $e$ is the sum of kinetic, thermal and
gravitational 
energies, $e=\rho (v^2+V)/2+P/(\gamma-1)$, and we have assumed an ideal
gas equation of state.  $V$ is the Newtonian gravitational potential
determined by Poisson's equation $\nabla^2V=4\pi G\rho$, $\rho$ is
the matter density, and $P$ is the pressure.

In terms of a flux vector $\vec{u}$ , we can write (\ref{eqn:euler}) as
\begin{equation}
\dot{\vec{u}}-\partial_i \vec{F}^i[\vec{u}]=0 ,
\label{eqn:vec}
\end{equation}
where $\vec{u}=(\rho,\rho v^1, \rho v^2, \rho v^3, e)$ is a five
component column vector, and  $\vec{F}^i$ is a $5\times 3$ matrix
function whose components can be read from (\ref {eqn:euler}).
We use the abbreviation $\partial_i \equiv \partial/\partial x^i$.

A general time dependent curvilinear coordinate transformation then
maps (\ref{eqn:vec}) into a new flux conservative system of equations
\begin{equation}
\partial_t (\rg u) + \partial_\alpha \left[ \rg e^\alpha_i(F^i-u
\dot{x}^i)  \right]= 0,
\label{eqn:moving}
\end{equation}
where $e^\alpha_i$ is the inverse triad ({\it dreibein}).  The
detailed derivation is shown in appendix \ref{app:curvi}.
It is given
as the inverse of the triad $e^i_\alpha=\partial x^i/\partial
\xi^\alpha$.  The volume element $\rg \equiv {\rm det}(e^i_\alpha)$ is
the determinant of the triad.  Note that the partial derivative for
time in equation (\ref{eqn:vec}) is performed by holding the Cartesian
coordinates constant, while the partial derivative for
time in equation (\ref{eqn:moving}) is obtained by keeping the
curvilinear coordinates constant.
We now need to specify the
differential coordinate transformation $\dot{x}\equiv \partial
x(\xi,t)/\partial t$
to close the system (\ref{eqn:moving}). 

As in Pen (1995), we define a coordinate transformation which is a pure
gradient
\begin{equation}
x^i=\xi^\mu\delta^i_\mu +\Delta x^i .
\end{equation}
where
\begin{equation}
\Delta x^i\equiv\frac{\partial \phi}{\partial \xi^\nu} \delta^{i\nu}
\label{eqn:ansatz}
\end{equation}
for some {\it deformation potential} $\phi$ to be defined later.  

\begin{figure}
\plotone{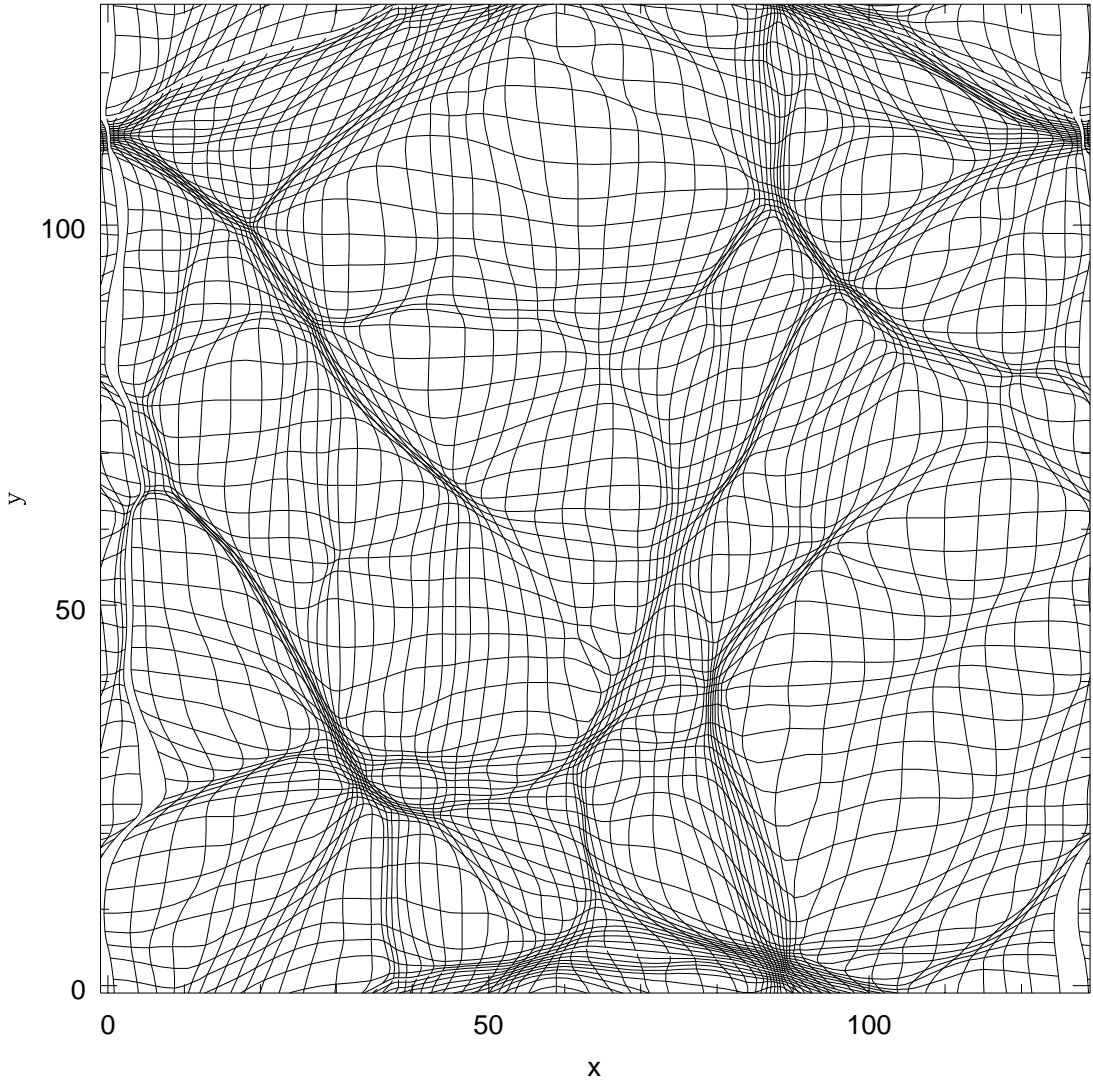}
\caption{A layer of the $128^3$ mesh of the CDM simulation projected
onto the $x-y$ plane.
For clarity, only every other grid line is plotted.  The salient
feature is the regularity of the grid.  Even in projection, the grid
never overlaps itself.  This is guaranteed by the compression limiters
since each curvilinear line
is a monotonically increasing function of its corresponding Cartesian
coordinate.}
\label{fig:mesh128}
\end{figure}

The triad is now explicitly symmetric
\begin{equation}
e^i_\alpha = \delta^i_\alpha + \phi_{,\alpha\beta}\delta^{\beta i}
\end{equation}
since partial derivatives commute.  In a
cosmological scenario, the initial conditions are almost smooth, and
we can set $\phi=0$ initially.  During subsequent evolution, we will
impose a constraint below to require a continuous sequence of
non-degenerate triads.  We are then assured that the  triad is
positive definite, from which it follows that 
\begin{equation}
\partial x^a/\partial\xi^a > 0
\label{eqn:pconstraint}
\end{equation}
(no summation).
We make several conclusions.  The triad has real eigenvalues, which
implies that the local coordinate transformation contains no rotation.
It is a triaxial locally conformal stretching of the curvilinear space
onto the Cartesian space.  From inequality (\ref{eqn:pconstraint}), it
follows that each Cartesian coordinate increases monotonically as a
function of its corresponding curvilinear coordinate.  So $x^1$ is
always monotonically increasing with $\xi^1$.  When projected down one
axis, the curvilinear maps never overlap themselves, as we indeed
observe in real simulations (see Figure \ref{fig:mesh128}).  We thus have a
mathematically rigorous formulation, where in the continuum limit
any triaxial object in the curvilinear coordinate system
aligned with the principal axes of the triad undergoes no rotation
when mapped into Cartesian space.  This is a
mathematical formulation of the statement that nearest neighbor
relationships are invariant of the deformation potential $\phi$.
Further properties of the curvilinear coordinate system are given in
appendix \ref{app:curvi}  Note that these results rely on the
implementation of a compression limiter described below.

The goal of astrophysical hydrodynamics has often been to maintain
constant resolution in mass coordinates.  The mass per unit
curvilinear coordinate volume is given by $\rg \rho=\rg u^0$, and its
evolution by the first component of equation (\ref{eqn:moving}):
\begin{equation}
\frac{\partial \rg \rho}{\partial t} + \frac {\partial}{\partial
\xi^\alpha} \left[\rg \rho e^\alpha_i (v^i-\frac{\partial
\dot{\phi}}{\partial \xi^\nu} \delta ^{\nu i}) \right] = 0
\label{eqn:ccons}
\end{equation}
If we desire the mass per volume element to be constant in time, we
set the first term in equation (\ref{eqn:ccons}) to zero and obtain
the linear elliptic evolution equation for the
deformation potential as in Pen (1995)
\begin{equation}
\partial_\mu \left( \rho \rg e^\mu_i \delta^{i\nu} \partial_\nu \dot{\phi}
\right) = \Sigma \equiv \partial_\mu \left( \rho\rg e^\mu_i v^i\right).
\label{eqn:def}
\end{equation}
We note that (\ref{eqn:def}) is linear in the deformation potential
$\dot{\phi}$, which resolves the dilemma that coordinate changes are
an inherently costly and nonlinear process.  The additional elliptic
equation increases the cost of the simulation by about a factor of
two, since we now have to solve two elliptic equations (the other
being for the gravitational potential) instead of one.  The evolution
of the deformation potential does not need to be very accurate
because the order of accuracy of the hydrodynamic calculation does not
depend on the choice of background geometry.

We solve for the gravitational potential using Poisson's equation
\begin{equation}
\partial_\mu( \rg g^{\mu\nu} \partial_\nu V) = 4\pi G
(\rho-\bar{\rho})\rg
\label{eqn:gravity}
\end{equation}
using the multigrid algorithm described in Pen (1995).  $\bar{\rho}$
is the mean comic density.

In a real simulation, we need to discretize the continuum equations.  In
order to maintain the good properties, we require that the local grid
is smooth, i.e. that the triad and therefore the deformation potential
do not change too much between adjacent cells.  This is achieved
through smoothing and compression limiters.

Smoothing is implemented by first smoothing the right hand side of
equation (\ref{eqn:def}) and then smoothing the time derivative of the
deformation potential before actually updating it.  We solve equation
(\ref{eqn:def}) for a smoothed divergence field, and smooth before
generating the deformation potential.  Furthermore, the current
implementation of the code uses an external multigrid routine, which
does not allow for a spatially variable mass function $\rho\rg$ in the
Laplacian operator of the deformation equation (\ref{eqn:def}), so the
current code simply drops that term.  This would cause a further
violation of the constant mass constraint for non-potential flow once
the constraint is violated.  Empirically, though, this effect is
insignificant, and is typically smaller than the contribution from the
limiters. In equation (\ref{eqn:limiter}) the current implementation
of $\Sigma$ does not contain $\rho\rg$, either.

To prevent excessive compression and the associated computational cost,
we add a compression limiter as described in Pen (1995).  In some
cases it is also desirable to introduce an expansion limiter to
maintain a minimal length resolution independent of density.  To
incorporate these crucial requirements, we introduce an auxilliary
variable $\Delta \phi$ from which $\phi$ will be derived as follows:
\begin{eqnarray}
\partial_\mu \left( e^\mu_i \delta^{i\nu} \partial_\nu \Delta{\phi}
\right) &=& S (\Sigma + C + E)\nonumber \\
\dot{\phi}&=& S \Delta\phi
\label{eqn:limiter} 
\end{eqnarray}
We define the compression limiter $C$ and expansion limiter $E$ as
\begin{eqnarray}
C(\phi)&\equiv& 4\left( \frac{\xi_m}{\lambda_0}-
	H(\frac{\xi_m}{\lambda_0}-1) \right)^2,
\nonumber \\
E(\phi,\Sigma) &\equiv& -2 H(\rg-v_m) |\Sigma|,
\label{eqn:clim}
\end{eqnarray}
where $H$ is the Heaviside function and $\xi_m
\approx 1/20$ is the maximal compression factor and $\lambda_0$ is the
minimum eigenvalue of the triad
$e_\mu^i$.  We choose a typical expansion volume limit $v_m=10$. The
smoothing operator $S$ is simplest to implement by 
smoothing over nearest neighbors in curvilinear coordinates.  We see
that the final deformation potential $\phi$ is always smooth on scales
which are smoothed by $S$.  We found  it empirically sufficient to
use a single Jacobi relaxation iteration in $\xi$ space for $S$.
Equations (\ref{eqn:limiter},\ref{eqn:clim}) differ from previous
implementations (FM, TF) by being locally defined.

This completes our description of the analytical formulation.

\section{Relaxing TVD}

One of the simplest high resolution TVD schemes to implement is the
relaxing TVD method (Xin and Jin 1994).  For completeness, the full
algorithm is described in appendix \ref{app:relaxing}  It has the
advantage of requiring no non-linear characteristic field decomposition
nor complex Riemann solvers.  Furthermore, it is not dimensionally
split, which is a desirable attribute in an algorithm such as ours
where the grid can become strongly skewed.  There is  no
need to explicitly evaluate the flux Jacobian eigenvectors.

We note that the flux limiter is applied to the hydrodynamic
quantities, but not the gravitational terms which are elliptic source
terms.

\subsection{Hydrodynamic Tests}

We now have a complete framework to test the adaptive mesh
hydrodynamics.  First, we test the accuracy of the relaxing TVD scheme
using the Sod shocktube test.  The test is performed as follows: we
start with a horizontal tube of gas and a membrane dividing the gas
into a chamber on the left and one on the right thereof.  The initial
state on the right is labeled using a subscript 1, and is defined by
some density and pressure $(\rho_1,p_1)$, and is taken to be at rest
with respect to the tube.  The state on the left of the membrane is
labeled using the subscript 4, and is given as $(\rho_4, p_4)$.  The
solution depends on the ratios of pressures and densities (Landau and
Lifshitz 1987 page 371), and we consider the case where $p_4>p_1, \
\rho_4>\rho_1$.  At an initial time $t_0$, the membrane is destroyed.
This results in a shockwave propagating into the the right side, whose
state we will describe using the subscript 2, and a rarefaction fan
penetrating the left side.  The initial discontinuity propagates
rightward, and we denote the region between the contact and the
rarefaction fan with a subscript 3.  We define the shock speed to be
$v_c$.  It follows that the velocity and pressure on both sides of the
contact are equal and constant $v_2=v_3, \ p_2=p_3$, and we solve for
the post shock pressure $p_2$ using the assumption of self-similarity
and the shock jump conditions
\begin{equation}
\frac{p_1}{p_4}=\frac{p_1}{p_2} \left[
1-\frac{\gamma-1}{2}\frac{c_1}{c_4}
(\frac{p_2}{p_1}-1)\sqrt{\frac{2/\gamma}{(\gamma+1)(p_2/p_1)+(\gamma-1)}}
\right]^{\frac{2\gamma}{\gamma-1}}.
\end{equation}
The sound speed
$c_i=\sqrt{\gamma p_i/\rho_i}$.  We can then solve for the remaining
quantities
\begin{eqnarray}
u_s &=& c_1
\sqrt{\frac{\gamma-1}{2\gamma}+\frac{\gamma+1}{2\gamma}
(\frac{p_2}{p_1})} \nonumber \\ u_2 &=& c_1 \left(
\frac{p_2}{p_1}-1\right) \sqrt{\frac{2/\gamma}{(\gamma+1)
(p_2/p_1)+(\gamma+1)}} \nonumber \\ \rho_2 &=& \rho_1
\frac{u_s}{u_s-u_2} \nonumber \\ \rho_3 &=& \rho_4 \left(
\frac{p_3}{p_4} \right)^{\frac{1}{\gamma}}.  \label{eqn:sod}
\end{eqnarray}

We choose the following parameters as initial conditions
\begin{equation}
\left( \begin{array}{c} \rho_1 \\
	p_1 \end{array} \right)
= \left(\begin{array}{c} 0.4 \\ 0.01  \end{array}\right) , \ \ \ \ \ \
\ 
\left( \begin{array}{c} \rho_4 \\
	p_4 \end{array} \right)
= \left( \begin{array}{c} 2 \\ 1 \end{array} \right)
\label{eqn:sodic}
\end{equation}
which are identical to the ones chosen in Shapiro et al (1995) to
allow for easy comparison.

\begin{figure}
\plotone{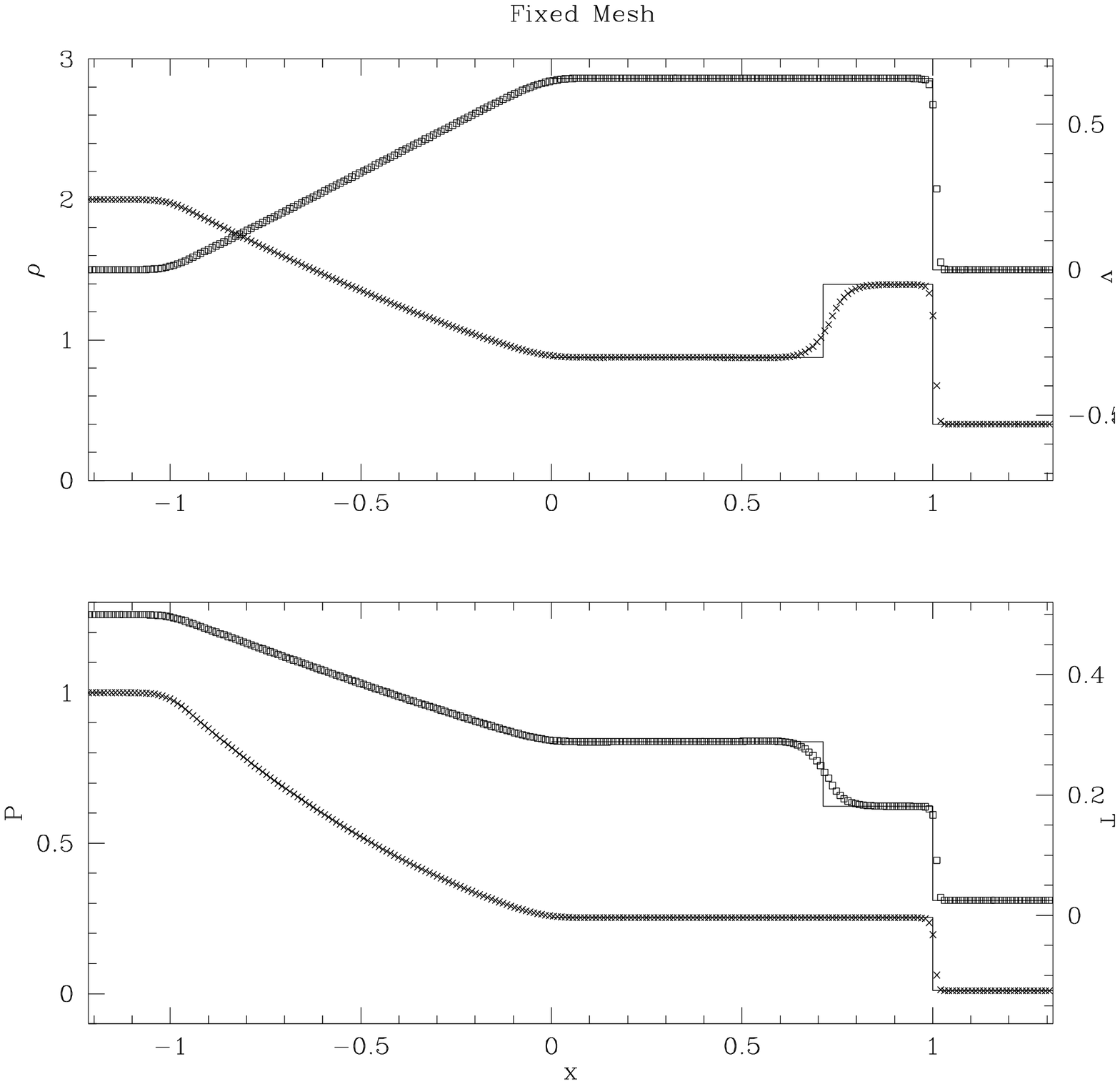}
\caption{Sod shock tube test on a fixed grid.  The crosses in the
upper plot are the numerically computed density points, and the boxes
are the numerical velocity field.  In the bottom graph, the crosses are
the pressure field, and the boxes are the temperature.  The solid lines
show the exact solution.}
\label{fig:sodfm}
\end{figure}

Figure \ref{fig:sodfm} shows the result of the Sod shocktube test
using the relaxing scheme in a fixed grid.  The plot has been rescaled
such that the shock position occurs at $x=1$.  To quantify the
resolution, we note that there are 98.38 cells between the initial
contact surface at $x=0$ and the shock front.  The relaxing scheme is
indeed well behaved, and provides non-oscillatory shock jump
conditions.  We also see that the contact surface has been
significantly diffused.  This is inevitable whenever one attempts to
advect a contact discontinuity for 70 cells across an Eulerian grid.
While some contact steepeners have been proposed in the literature
(Harten 1983), they cannot restore information which has been
inherently lost.  They work well when all contacts are well resolved,
but suffer from problems when either the time steps are short or if
more dynamical processes occur.  Furthermore, since contact
discontinuities are not evolutionary and arise only from singular
initial data, we should consider it safe to ignore any diffusion
across such a surface.

The shock front itself is accurately resolved within two cells, which
is comparable to most modern flux conservative hydrodynamic shock
capturing schemes.  While the classical diffusion schemes trade off
shock width against post shock oscillations and stability, the TVD
schemes have no free parameters for artificial viscosity.  We also see
the correct complete absence of oscillations about the contact
discontinuity.

\begin{figure}
\plotone{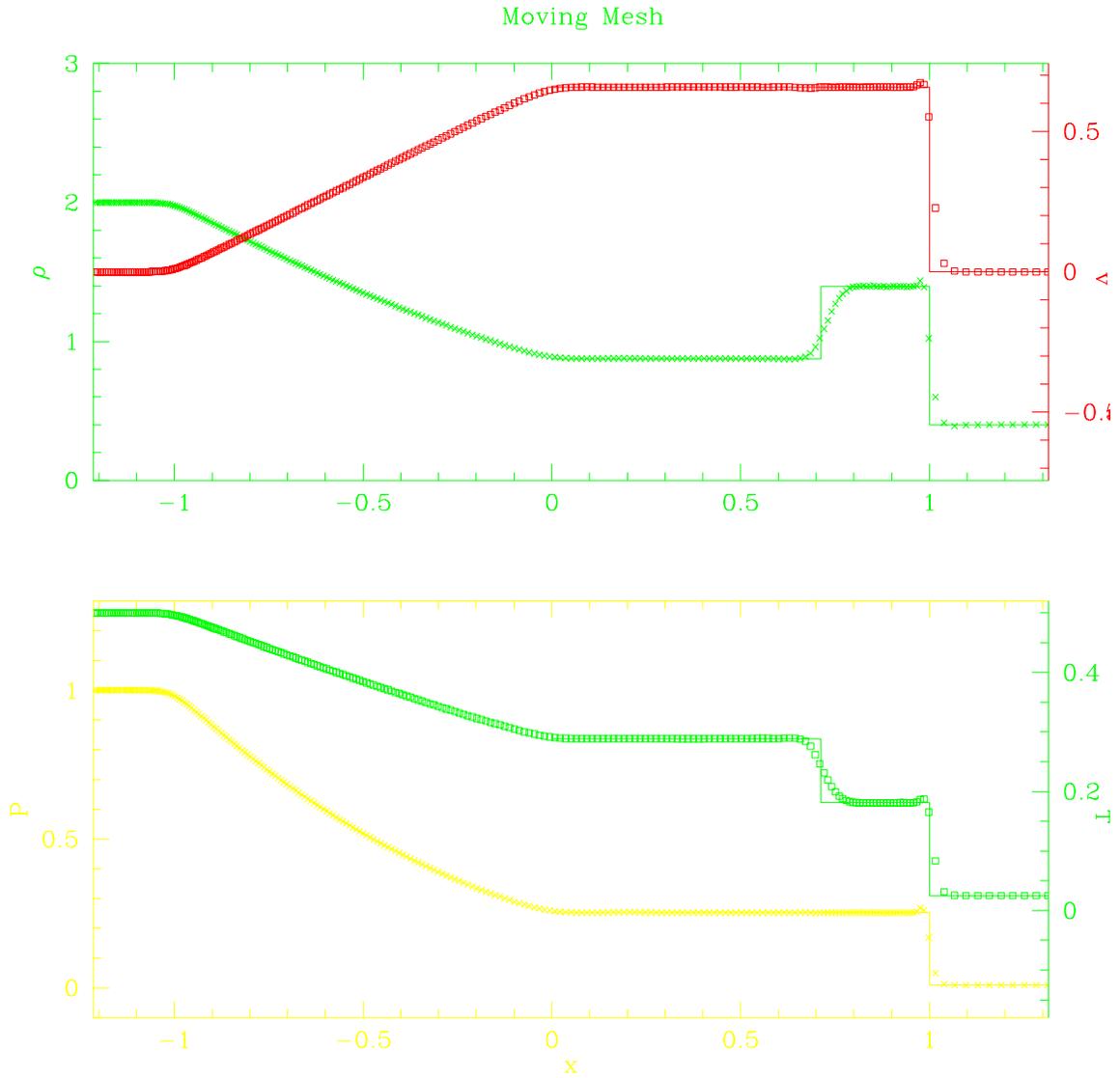}
\caption{Sod shock tube test on the moving mesh.  The notation is
identical to that of 
Figure \protect\ref{fig:sodfm}.  We see a slight postshock fluctuation
from the curvlinear transformation
explained in the text.}
\label{fig:sodmm}
\end{figure}

We now examine the moving in mesh one dimension.  Figure
\ref{fig:sodmm} shows the same shock tube problem given by formula
(\ref{eqn:sodic}) run using the full three dimensional moving mesh
code with slab symmetry.  The mesh is chosen with approximately
constant mass per grid cell, and in this run we have approximately 83
grid cells between the initial membrane $(x=0)$ and the shock front
$(x=1)$.  The solution is still well behaved, and the shock front is
also resolved in two cells.  We see a little overshoot just after the
shock front.  TVD is applied to the curvilinear characteristic fields,
and the solution has no overshoot when plotted in curvilinear
coordinates.  In the Cartesian frame is can appear as if overshoots
did form.  This could be circumvented by transforming to a Cartesian
frame before applying the limiters.  But the motivation behind MMH was
the success of SPH in tracking physics on constant mass scales, we
argue that applying TVD in curvlinear frames might even be physically
better motivated than its application in Cartesian space.  The moving
mesh relaxing TVD indeed appears to be a viable and accurate algorithm
at least for these rather trivial test samples.

A much more challenging and comprehensive test of the (gravity free)
moving mesh hydrodynamic code is a Sedov Taylor blast solution.  It
requires a large dynamic range since the exact solution piles up most
of material just behind the shock front.  We set up a box with
constant density and a large supply of thermal energy $E_0$ in the
center at $t=0$.  As the solution evolves, it tends towards the
self-similar Sedov Taylor solution (Shu 1992).  The evolution of the
shock radius is $R(t)=\beta(E_0 t^2/\rho_1)^{1/5}$ (Landau and
Lifshitz 1987 page 404), where $\beta \approx 1.15$ for a $\gamma=5/3$
gas.  In our test case we choose the ambient density $\rho_1=1$, and
$E_0=44577$.  The outside pressure is $10^{-3}$, which is our
numerical approximation to $0$.  

\begin{figure}
\plotone{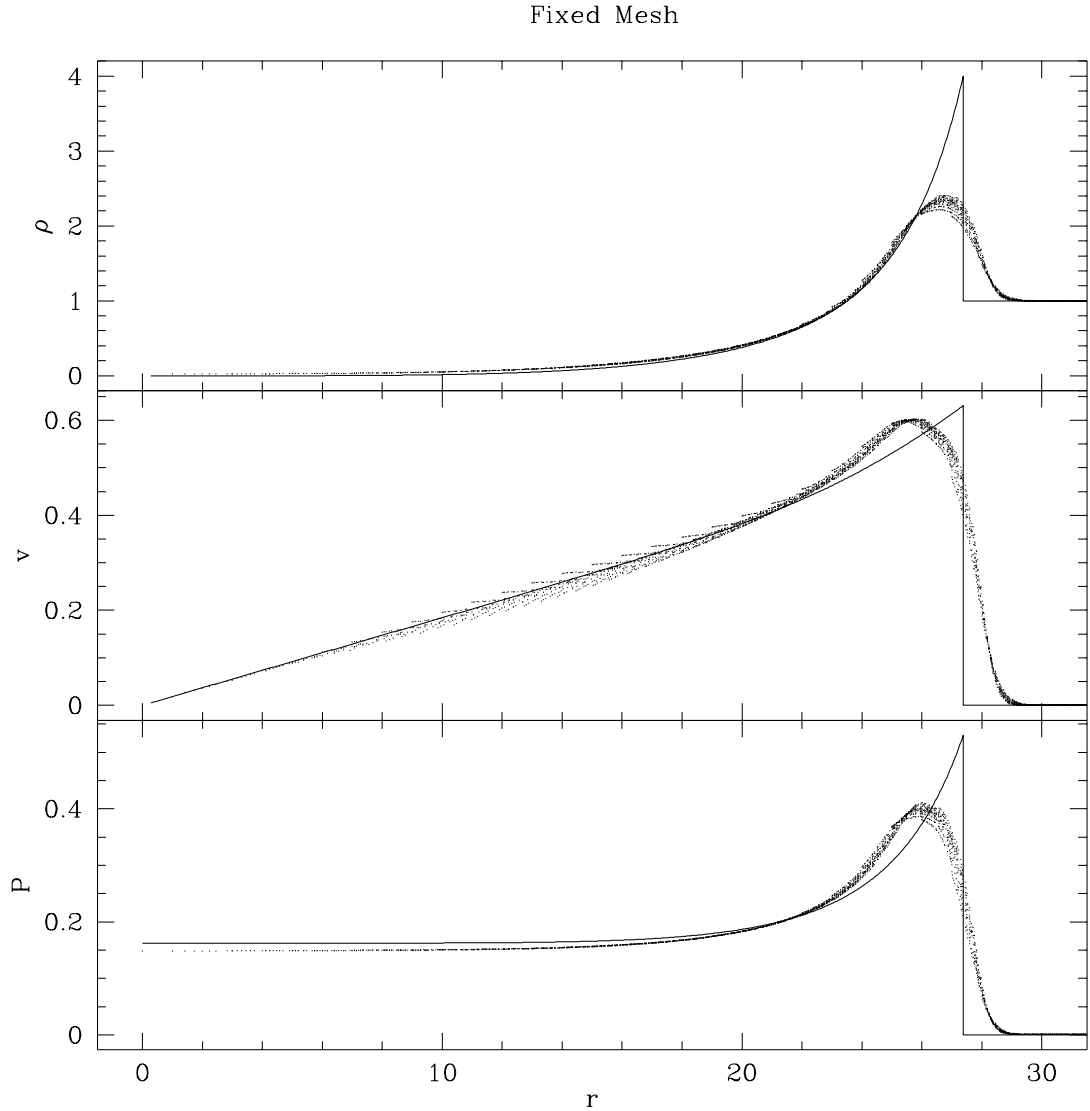}
\caption{Sedov Taylor explosion with energy input $E=44577$ at $t=0$
shown at time $t_f=13$.  The grid points are projected along all
angles, and 
the mesh contains one grid space per unit distance, with one point per
computational grid cell.  The shock widths
are about two grid cells.  The solid line is the exact solution.}
\label{fig:sedovfm}
\end{figure}

\begin{figure}
\plotone{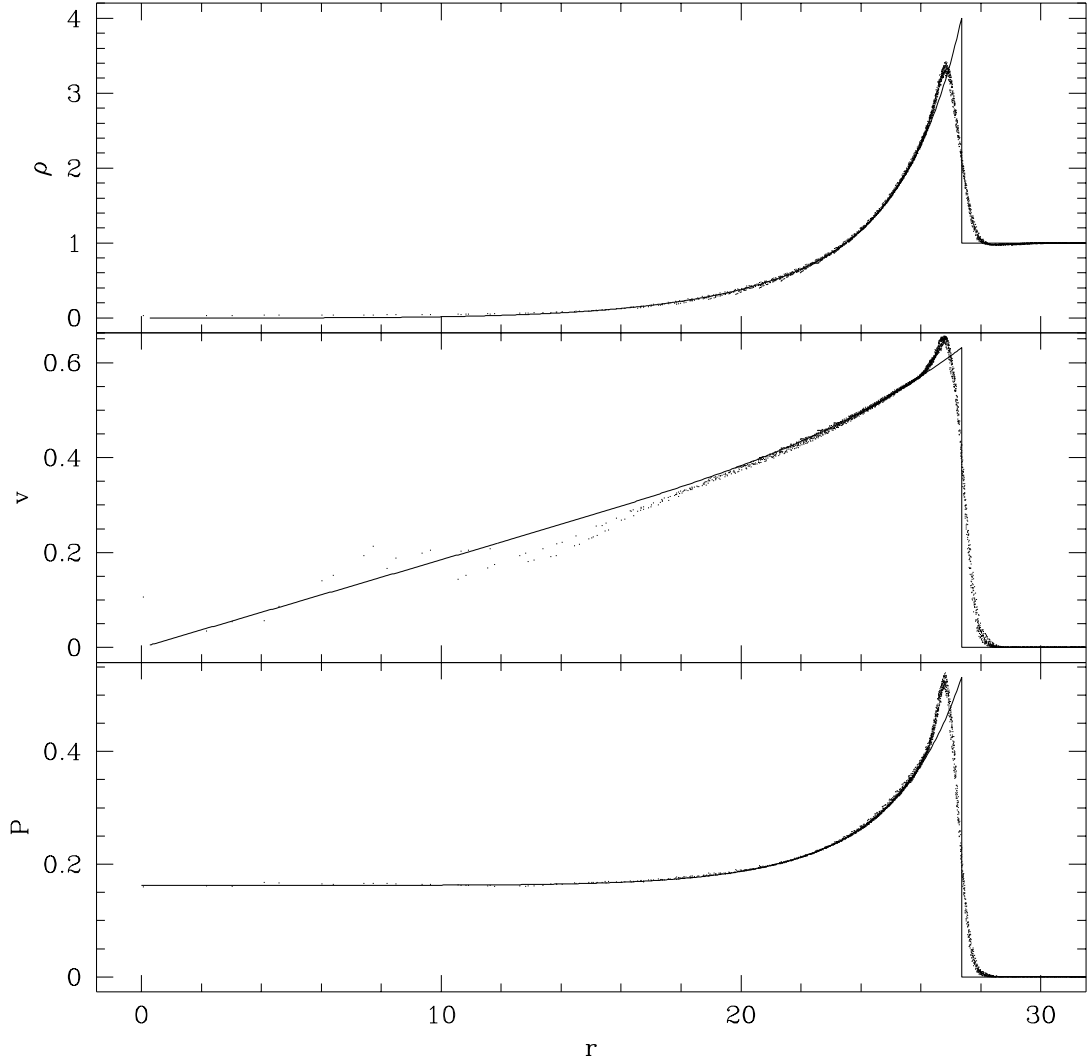}
\caption{Same as Figure \protect\ref{fig:sedovfm} but on the moving
mesh.  Angular anisotropy is reflected in the width of the line, which
we see has decreased compared to the fixed mesh.  The
shock is not less than one half Cartesian unit wide.  Since the mesh
compresses by at most four times across the shock, the average density
is two, and we would expect twice
the shock resolution of figure \protect\ref{fig:sedovfm}.  
}
\label{fig:sedovmm}
\end{figure}

\begin{figure}
\plotone{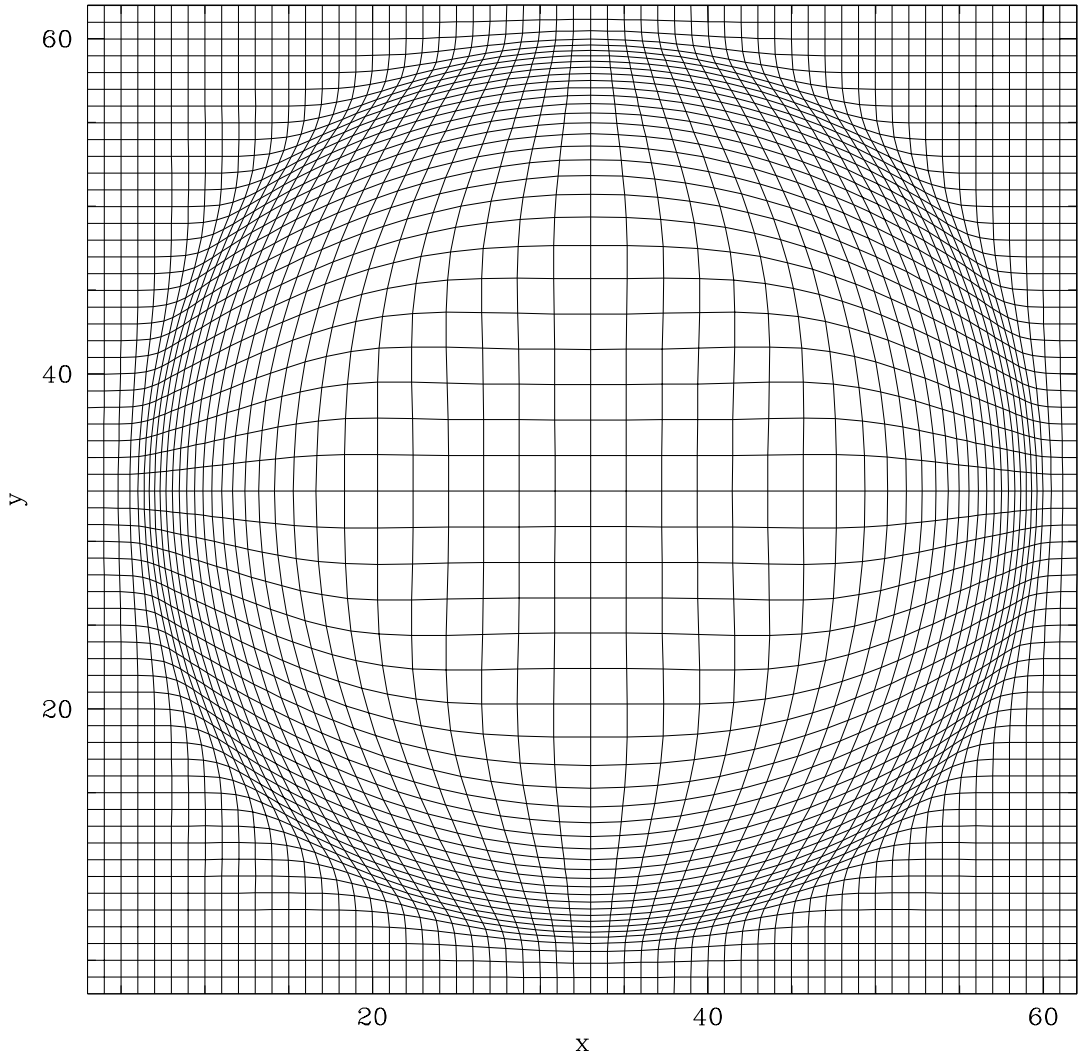}
\caption{Mesh geometry at midplane for Sedov Taylor simulation.
The
expansion limiter prevents the cells from expanding more than a factor
of 10 in volume.}
\label{fig:sedovmesh}
\end{figure}

In figure \ref{fig:sedovfm} we show the full three dimensional solution
on a fixed mesh projected onto the radial coordinate.  Each grid cell
is plotted as one point.  The scatter is due to the anisotropy of the
Cartesian grid, which occurs since the thin shock layer is not fully
resolved, and the resolution is a function of the angular coordinate.
The resolution is necessarily lower along diagonal directions.  At this
resolution we see that even the shock jump condition, which would imply
a postshock density of 4, is not well resolved.  This  implies that the
shock amplitude will be a strong function of resolution.  When we
examine the performance on the moving mesh in figure \ref{fig:sedovmm},
we see that the mesh just postshock compresses by a factor of 4, raising
the resolution by that amount.  In the interior, however, the mesh
expands drastically, and we significantly degrade resolution, as can be
seen by the scatter at smaller radii.  Figure \ref{fig:sedovmesh} shows
the mesh at the end of the blast wave.

\section{Cosmological Hydrodynamics}

We can preserve the exact time invariant conservation form of the fluid
equations in a Friedman-Robertson-Walker (FRW) expanding background by using
the expansion changing to comoving variables $q=ax$ and a new time
scale (Gnedin 1995)
\begin{equation}
d\tau=\frac{dt}{a^2}.
\label{eqn:newton}
\end{equation}
Using this variable, Newton's laws apply
directly, and in particular objects travel on straight trajectories
unless acted upon by another force.  The cost comes as a time
dependence of Newton's constant.  We will call the new time coordinate
$\tau$ the {\it Newtonian} time frame.  To further fix our units, we
define the scale factor today $a_0=1$.

The scale factor $a=t^{2/3}$ is given in a flat universe as
\begin{equation}
a=\frac{9}{\tau^2}
\end{equation}
where $-\infty<\tau<0$, and the proper time $t=-8/\tau^3$.  In a
curved universe,
\begin{equation}
a=\frac{9}{\tau^2+9\kappa}
\label{eqn:curv}
\end{equation}
where $\kappa=(\Omega_0-1)/\Omega_0$ is related to the
curvature scale.  Again, we obtain an abrupt end to the Newtonian time
for a hyperbolic universe, where the gravitational interaction becomes
infinitely strong at $\tau=\sqrt{-\kappa}/3$.  More curious is the
fact that in a closed universe, (\ref{eqn:curv}) the Newtonian time
extends across the full  real number line, and in fact the turnaround
occurs at $\tau=0$, after which the gravitational interaction weakens
again.

Unfortunately the case with any cosmological constant has no exact
solution for the scale factor $a$, and we integrate the Friedman
equation
\begin{equation}
\left(\frac{da}{d\tau}\right)^2 = \frac{4a^3}{9} \left[ 1 + a^3
\Omega_\Lambda/\Omega_0 -a \frac{\Omega_0+\Omega_\Lambda-1}{\Omega_0} \right]
\end{equation}
to third order in the Taylor expansion at each time step.  The various
values of $\Omega_i$ are given at today's epoch where $a_0=1$.

We can qualitatively understand such a universe with small
initial perturbations.  Initially the perturbations grow, become
non-linear,  shock heat and form into clusters, pancakes and
filaments.  This process stops as one approaches turnaround at $\tau=0$.
After turnaround, the physical processes are dominated by
hydrodynamic interactions, with gravity becoming less and less
important in the evolution of the gas, which redistributes itself into
pressure equilibrium.  The final distribution is determined by the gas
entropy distribution at turnaround, with the gas at low entropy
condensing into high density regions, and gas at high entropy
distributed tenuously spread over a larger volume.  Paradoxically, the
regions of low entropy are the voids at turnaround, which will become
high density regions.  The cluster outskirts have the high entropy,
and will fill most of space.  The cluster cores are in between, and
will expand from the compressed configuration.

Using the Newtonian time $\tau$, Equation (\ref{eqn:euler}) maintains the
identical hydrodynamic interaction, but with a time dependent
gravitational source term
\begin{eqnarray}
\frac{\partial\rho}{\partial \tau} + \frac{\partial}{\partial x^i}
\rho v^i &=&0 
\nonumber\\
\frac{\partial\rho v^i}{\partial \tau} + \frac{\partial}{\partial x^j}
\left[ \frac{}{}  \rho v^i v^j + P \delta^{ij} \right.
\ \ \ \ \ &&\nonumber \\
\left. - a\bar{\rho} V+\frac{a}{4\pi G}\left(
\frac{\partial V}{\partial x^k} \frac{\partial V}{\partial
x^l}(\delta^{il} \delta^{jk}-\frac{1}{2}\delta^{ij}\delta^{kl}) \right)
\right]
 &=&0 
\nonumber\\
\frac{\partial e}{\partial \tau} + \frac{\partial}{\partial x^i}\left[
(e+\rho V+P) v^i \right] &=& -a\rho v^i V_{,i} \nonumber \\
\nabla^2 V &=& 4\pi G (\rho-\bar{\rho}).
\label{eqn:cosmo}
\end{eqnarray}

In a numerical code, we have several choices about units.  In order to
keep quantities close to unity, we use units where the grid spacing
$\Delta x_g=1$, which defines the conversion factor for length $x_l=L
x_g$. We further simplify our units by choosing $6\pi G\equiv 1$.
For density, we define the mean density of a fluid to be one,
so the average mass per cell for each fluid is $<\rho\rg>=1$.  This
fixes the mass unit $m_l=M m_g$ in terms of the critical density
$M=\Omega_b \rho_c L^3$ where $\rho_c=3H_0^2/8\pi G$.  $\Omega_b$ is
the gas fraction in units of the critical density.  The time unit has
already been completely fixed, and is given by $t_l=T t_g$ where
\begin{equation}
T = \sqrt{\frac{\Omega_bL^3}{6\pi G\Omega_0 M}}.
\end{equation}

The comoving quantities (subscript $c$) are related to the lab
values (subscript $l$) by the scaling
\begin{eqnarray}
\rho_c &=& a^3\rho_l \\
\bar{\rho} &=& \frac{1}{6\pi G} \label{eqn:rhoh}\\
v_c &=& a(v_l-v_h) \\
v_h &=& \frac{\dot{a}}{a} x_l \\
e_c &=& a^5 e_l \\
V_c  &=& a V_l \\
x_c &=& a x_l \\
P_c  &=& a^5 P_l 
\end{eqnarray}
We note that the exact conservation of energy is lost, while momentum
is still conserved.  The latter is retained since FRW maintains space
translation invariance, but time translation invariance has been
explicitly destroyed.

\subsection{Energy Conservation}

In the presence of gravity in an expanding universe, the hydrodynamic
energy $e$ in equation (\ref{eqn:cosmo}) has a gravitational source
term.  By integrating the energy equation 
in (\ref{eqn:cosmo}) over space and time, and applying the continuity
equation, we obtain the Layzer-Irvine equation (Peebles 1981)
\begin{equation}
e(t_f) + g(t_f)= e(t_i)+g(t_i)
-a(t_f)\int_{t_i}^{t_f} \frac{e(t)}{a(t)^2} dt
\label{eqn:li}
\end{equation}
where $g\equiv a\int \rho V d^3x/2$ is the gravitational binding
energy.  We see that the sum of potential and kinetic energies is
negative, with the source term being the path dependent quantity under
the integral sign in (\ref{eqn:li}).  Typically this path dependent
term contributes 20\% of the
magnitude of the potential energy.  We define the dimensionless
Layzer-Irvine energy conservation ratio as in Ryu et al (1993)
\begin{equation}
R\equiv
\frac{-(e(t_f)-e(t_i)-g(t_i) + a(t_f)\int_{t_i}^{t_f} \frac{e(t)}{a(t)^2}
dt)}{g(t_f)}.
\label{eqn:lir}
\end{equation}
This quantity should be unity if energy is exactly conserved.
Throughout the run we can monitor this quantity $R$,
which gives us some indication about the errors in the simulation.
For the CDM power spectrum (Bardeen \etal\ 1986), where most of the
power is at the grid 
scale, an evolution of the test case spectrum results in $R\approx
1.3$, which implies a substantial energy error.  This error is easily
understood since numerical diffusion always smoothes out the density
field, thereby lowering the magnitude of the gravitational binding
energy.  The error decreases to $R \approx 1.1$ when we compute on a
moving mesh.  Since the primary contribution of power comes from small
scales, and since the grid is smoothed, the moving mesh cannot in fact
resolve the diffusion problem arising on the small scales where the
grid does not follow the fluid at all.

For grid based schemes, a significant source of error arises due to
artificial diffusion.  Even though the TVD schemes in principle have no
explicit diffusion or viscosity, the TVD limiter upwind modifies the
mass flux (\ref{eqn:ccons}).  The moving mesh reduces the mass flux
over a Cartesian Eulerian grid, and in principle the mass flux is
identically zero for potential flows.  In this case, the limiter
introduces no diffusion at all.  In practice, though, the grid
compression limiter (\ref{eqn:clim}) causes the grid to break away
from the fluid flow.  Furthermore, the grid only tracks the fluid to
first order accuracy, which leads to some variation in mass
(typically a few percent).  Another major effect is the generation of
vorticity.  In a pure vorticity equilibrium with no potential flow,
the mass in each volume element is constant, but each of the
directional fluxes is nonzero.  Only the sum is zero.  In this case,
the flux limiter will kick in, which attempts to reduce extrema.

Since the energy conservation, or in our case the Layzer-Irvine energy
(\ref{eqn:lir}), is a global quantity, we need to look for local error
estimators in order to assess the uncertainty in the physical
observable quantities such as X-ray luminosity, mass functions, etc.
For this purpose, we can add a term in the equation of motion which
has the same magnitude as the energy error, from which we can gauge
the propagation of errors.  Instead of funneling all the diffusion
error into the gravitational binding energy, we divert the entirety
into thermal energy errors.  The physical interpretation would be as
follows: A gravitationally bound object, say a cluster of galaxies,
moves through the grid.  As a result of the motion, the cluster
diffuses, resulting in an increase in its core radius.  This decreases
the gravitational binding energy, and thereby violates the virial
theorem.  The cluster expands even further to reach a new equilibrium.
The alternative scenario would be to decrease the thermal energy at
the same time.  The thermal energy is the only Galilean invariant that
we can use, since kinetic energy depends on the frame.  Furthermore we
know that momentum is fundamentally conserved, and since the main
purpose of cosmological gas dynamics is to accurately calculate the
difference between gas and dark matter, we want to resist touching the
momentum equation.

The limiter in the continuity equation (\ref{eqn:ccons}) can be
represented as a diffusive flux vector $D^\alpha$ such that the
continuity equation becomes
\begin{equation}
\partial_t (\rho\rg) + \partial_\alpha \left[ D^\alpha +  e^\alpha_i
\rho \rg(v^i-\Delta \dot{x}^i ) \right] = 0.
\end{equation}
When we integrate the energy equation over space, we obtain a source
term as a function of $D^\alpha$
\begin{eqnarray}
\partial_t(\int e\rg d^3\xi) &=& - a \int \rho v^i e^\alpha_i
V_{,\alpha} \rg d^3\xi 
\nonumber \\
&=& a \int \left[ V \partial_t(\rho \rg) - V D^\alpha_{,\alpha}
\right] d^3\xi.
\label{eqn:esource}
\end{eqnarray}
We can therefore add the (positive value) of the second term in
(\ref{eqn:esource}) to the right hand side of the energy equation in
(\ref{eqn:cosmo}) to cancel its effect in (\ref{eqn:esource}).  With
this prescription, the only source for energy errors are due to time
discretization, and we indeed observe that $R=1$ for short time steps
using a fixed grid.
For the cosmological runs at maximal Courant time step, the error in
$R$ is typically a few percent.

\begin{figure}
\plotone{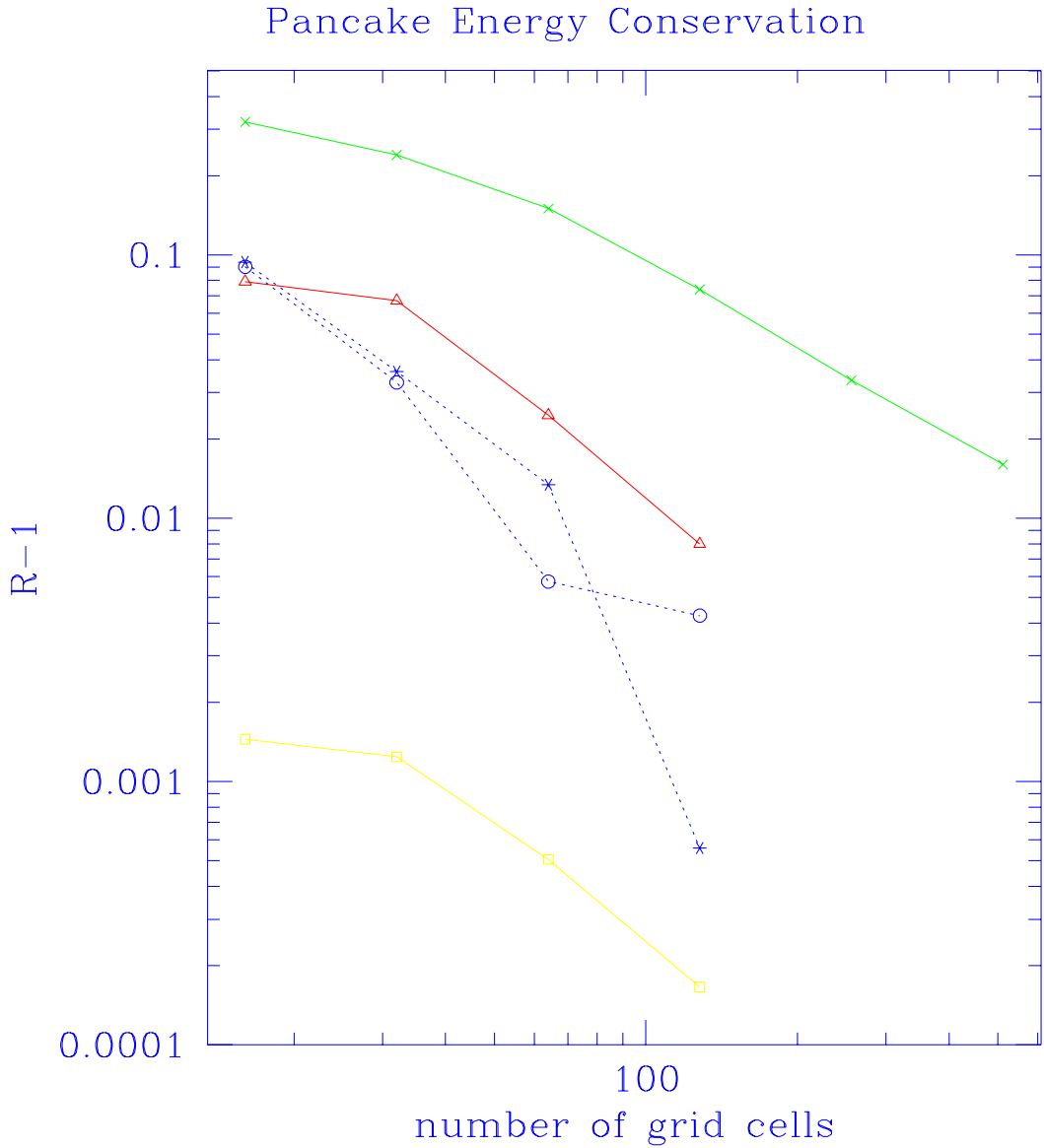}
\caption{Layzer-Irvine energy error for the pancake test.  The
vertical coordinate $R$ is defined in the text, and is
proportional to 
the kinetic energy divided by the potential energy.  Solid lines are
for fixed mesh calculations.  Open symbols have the energy
compensation scheme built in.  The boxed symbols on the bottom solid
line are run using a fixed
mesh with energy compensation and a time step which is $1/50$th of
the usual time step.}
\label{fig:penergy}
\end{figure}

We measured the energy error using each of these two schemes for the
pancake test described below.  The result is shown in Figure
\ref{fig:penergy}.  Since the parameter $R$ in equation
(\ref{eqn:lir}) contains kinetic energy divided by potential energy,
we would expect diffusion to always increase $R$, which is indeed
observed.  The top line with crosses shows the energy error for the
standard fixed mesh code without any corrections.  Since the diffusion
error is first order in space due to upwind limiting, we expect the
error to decrease linearly with resolution, as indeed it does.  Even
when the time step is reduced by a factor of 50, that error
changes by less than 10\%.  Time discretization does not contribute
significantly.  The story changes when we implement the energy
compensator (\ref{eqn:esource}).  The solid line with open triangles
shows an immediate decrease in the energy error.  The error now arises
primarily from time discretization, and by reducing the time step by a
factor of $50$ in the bottom line with open squares, the error also
decreases by that amount.  Our current energy compensation scheme is
first order accurate in time.  On a moving mesh, the energy
compensation has little or no effect, as we can see from the dotted
lines.  Energy is already much better conserved since the mass fluxes
and therefore the limiter diffusion terms are significantly smaller.

When we perform a run both with and without the source term in
(\ref{eqn:esource}), we compare the thermal energies at the end of the
run, from which we learn which cells have a large error and which ones
don't.  This is demonstrated in the pancake test described below.
Empirically we find that energy conservation is always good whenever
the power is well resolved in mass units.

\subsection{Time Step}

We have three factors which determine the time step, and choose the
smallest of the three.  Firstly, we have the Courant condition, which
requires that the maximal characteristic travels less than
$1/\sqrt{3}$ grid cells in one time step for a three dimensional
unsplit code.  In practice, we choose $t_{cfl}$ as half of that value.
For cold or high Mach number flows, the moving coordinate system
lowers the characteristic speeds in the curvilinear frame, where at
zero temperature and potential flow the characteristics would be
stationary in the curvilinear frame.  Nevertheless, there is still a
time step constraint, which is related to the divergence of the
velocity field.  It has dimensions of inverse time, and we define
$t_z$ as $1/8$th of the inverse of the smallest eigenvalue of the
matrix $\partial v^i/\partial x^j$.  The last time scale is determined
by the cosmic expansion.  We require that $\Delta a/a <
1/50$ between gravitational time steps, thus setting $t_c$.

In practice, a simulation is always dominated by the cosmological
expansion $t_c$ initially, but most of the CPU time is spent in the final
non-linear clustering stages where $t_{cfl}$ and $t_z$ are typically
closely balanced.

\subsection{Cosmological Tests}

\begin{figure}
\plotone{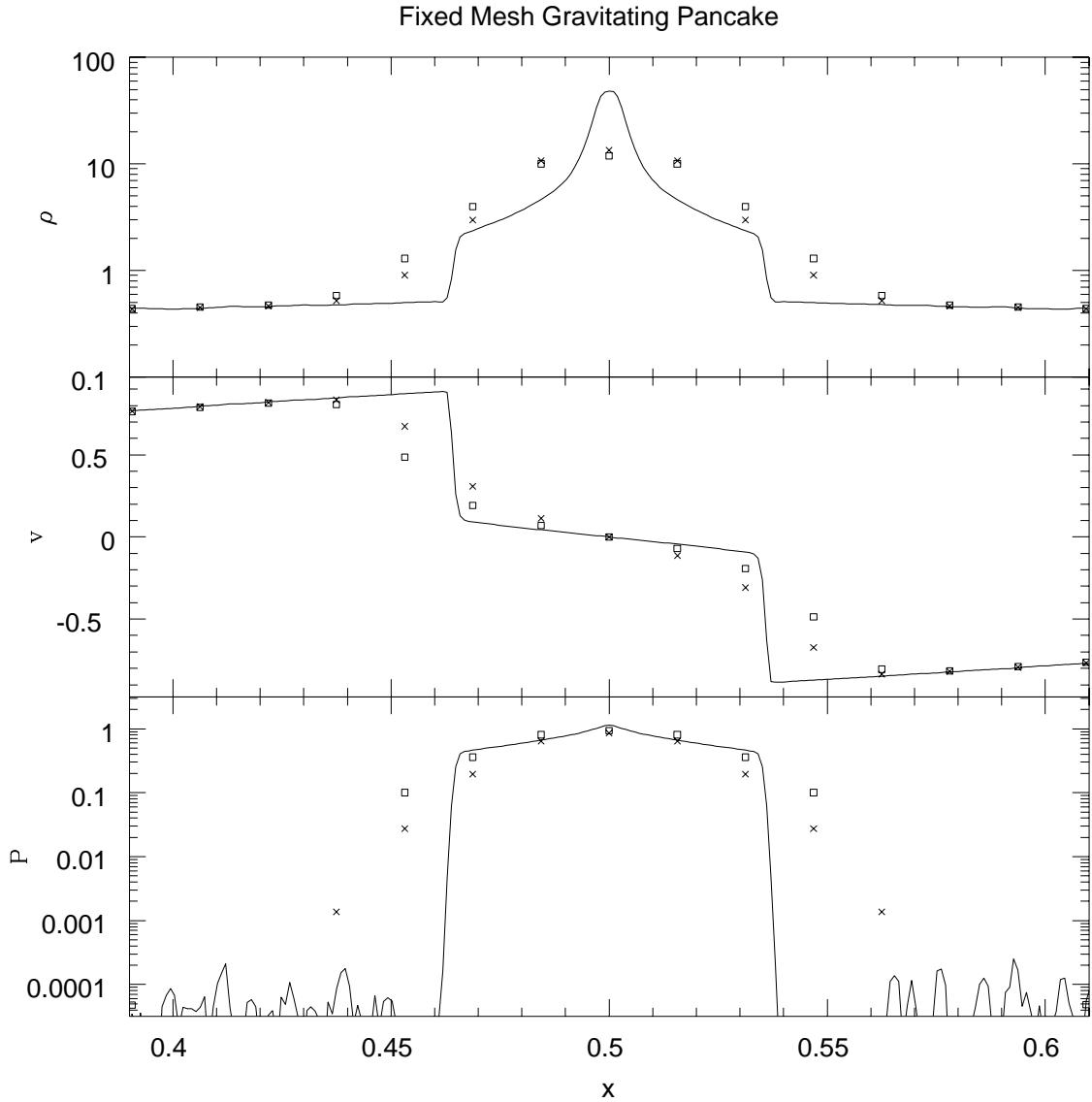}
\caption{The cosmological pancake test on a fixed mesh with 64 grid
points.  The open squares are run
without energy compensation, while the crosses have both energy
compensation and a shortened time step.  The solid line is the
solution obtained on a grid with 1024 points.}
\label{fig:pfm}
\end{figure}

\begin{figure}
\plotone{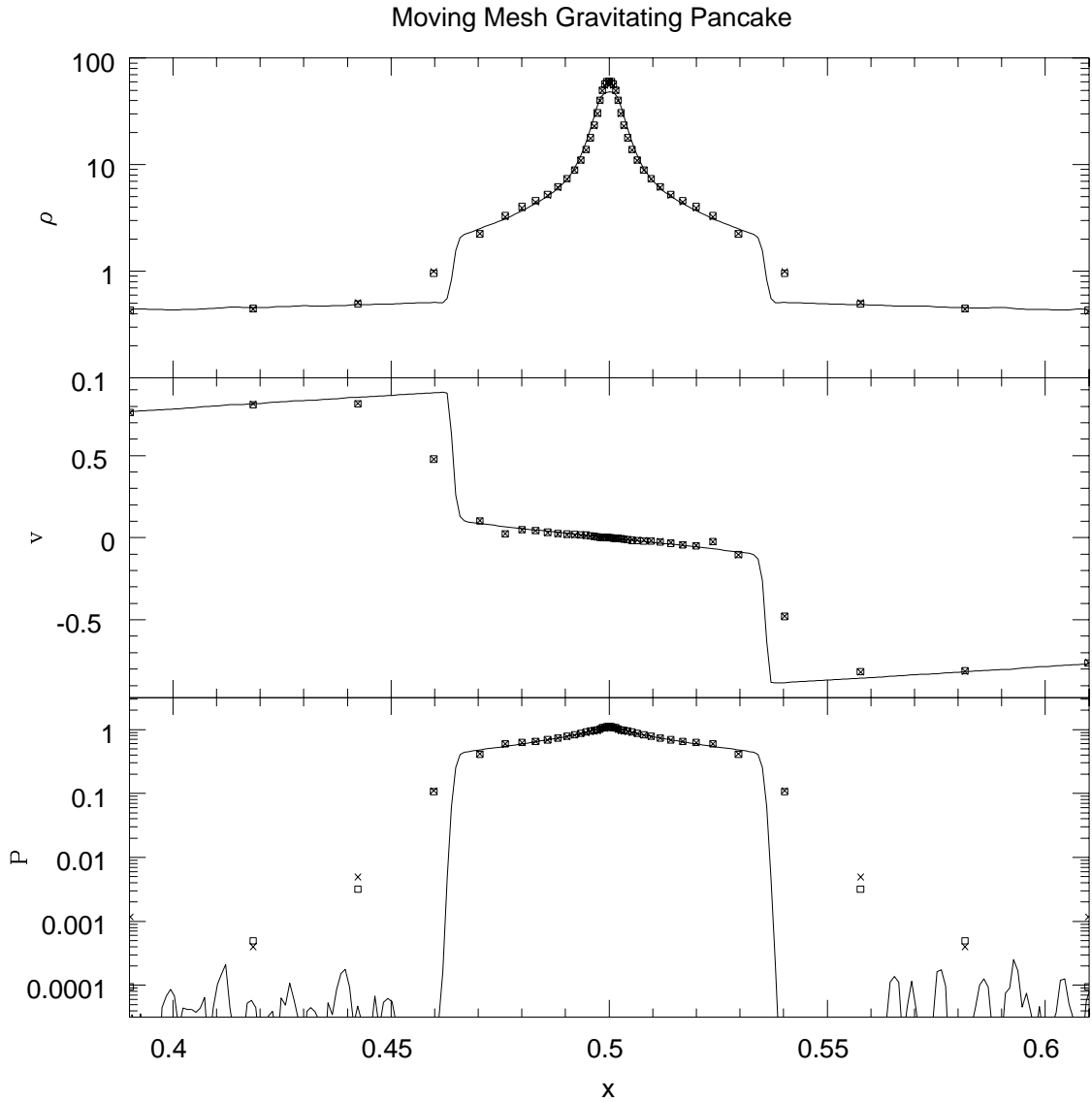}
\caption{The cosmological pancake test on a moving mesh with 64 grid
points.  The open squares are run
without energy compensation, while the crosses have energy
compensation using the same time step.  The solid line is the
solution obtained on a grid with 1024 points.  On a moving mesh, the
energy compensation does not make a significant difference.  The
maximal limiter $\xi_m=1/30$, using which the moving mesh achieves a
higher central density and resolution than the 1024 fixed mesh.}
\label{fig:pmm}
\end{figure}

The gravitational pancake as described in Ryu et al (1993) was
tested.  We set up a convergent wave at $z=21$ which collapses at
$z=1$.  The test was run on a fixed grid of $1024$ cells, and then
using fixed and moving meshes on $64$ cells.  In Figure \ref{fig:pfm}
we see an under resolved pancake on the fixed mesh.  For such a
structure, the effect of energy compensation is significant, and the
discrepancy between the compensated and uncompensated energy solutions
gives a good estimate of the mass diffusion and gravitational error.
When we run the simulation on a moving mesh in Figure \ref{fig:pmm}, 
we obtain a much better resolution of the pancake core, if we use the
central density as an indicator of resolution.  In fact,
using the compression limiter $\xi_m=1/30$, the $64$ cell moving mesh
outperforms the $1024$ cell fixed mesh.  We further notice that the
energy compensation has only a small effect on the solution, which
again is due to the fact that mass diffusion is a lesser problem in
our moving coordinates.

\begin{figure}
\plotone{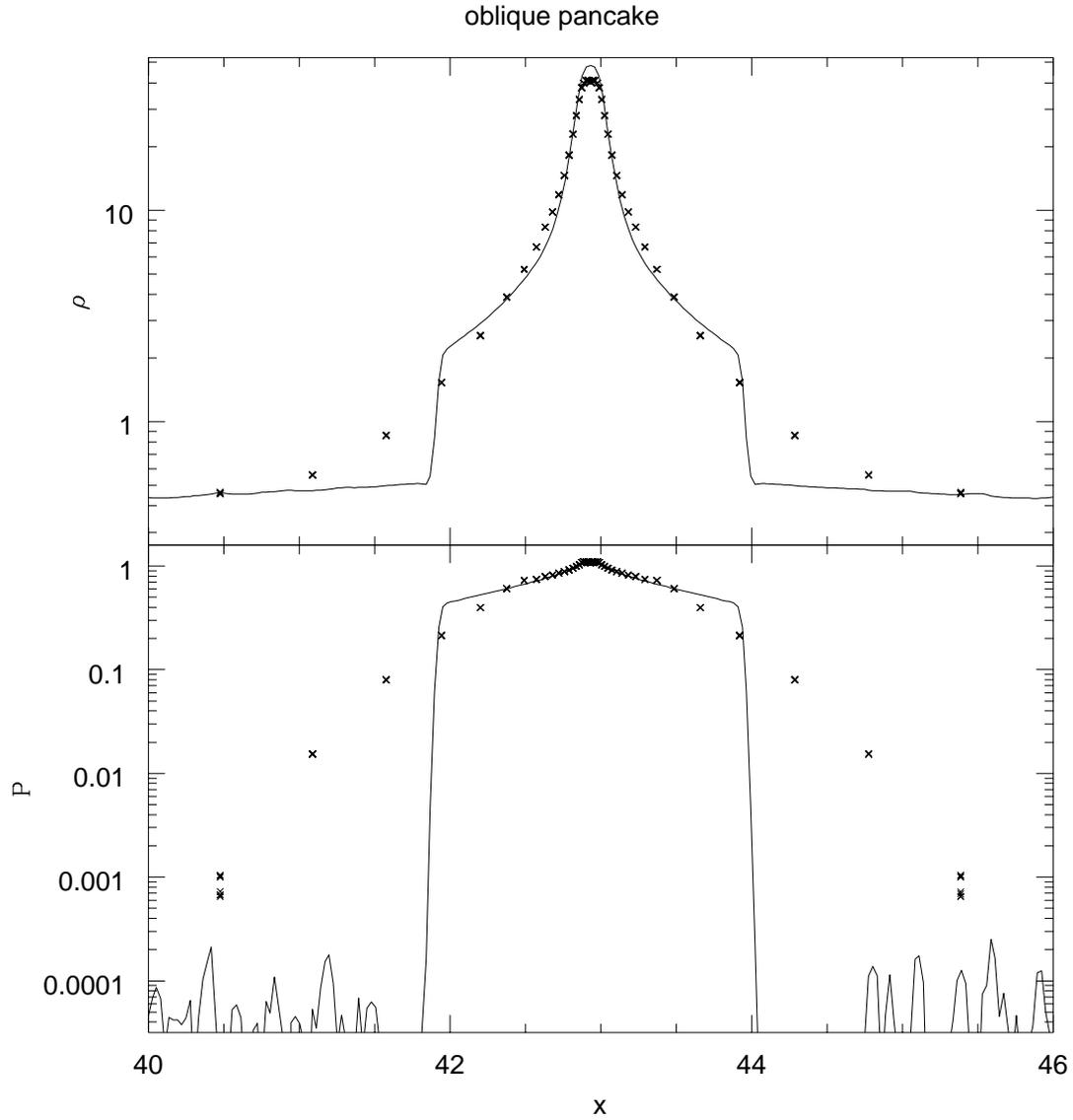}
\caption{
A pancake aligned on an $\tan^{-1}(1/2)$ angle to a $64^2$ grid.  The
horizontal axis is on grid units.  The crosses represent grid cells
rotated into the plane perpendicular to the pancake.  The solid line
is the rescaled solution from the 1024 cell fixed mesh.  Due to
projection effects, the spacing between cells which appear
adjacent in projection is $1/\protect\sqrt{5}$ of the actual
perpendicular nearest neighbor distance.  If we estimate the
shock width as 
four cross spacings, we would have an effective shock width and
resolution of 1.8 grid units.
}
\label{fig:oblique}
\end{figure}

An intriguing challenge is to collapse a pancake along a diagonal axis
in two dimensions at an angle $\theta=\tan^{-1}(1/2)$.  The path of
each vertex in the central pancake region is such that it does not
intersect its nearest neighbor, nor any of the diagonal neighbors.
This tests the code with a strong shock in a highly distorted and
oblique geometry.  Note that in this configuration the discretized
elliptic equations are no longer diagonally dominant, which is a
further test for the potential solvers.  We set up a run using a
$64^2$ mesh with a maximal compression limiter $\xi_m=1/30$.  Despite
these challenges, the code performs optimally for the high density
regions as shown in figure
\ref{fig:oblique}.

\begin{figure}
\plotone{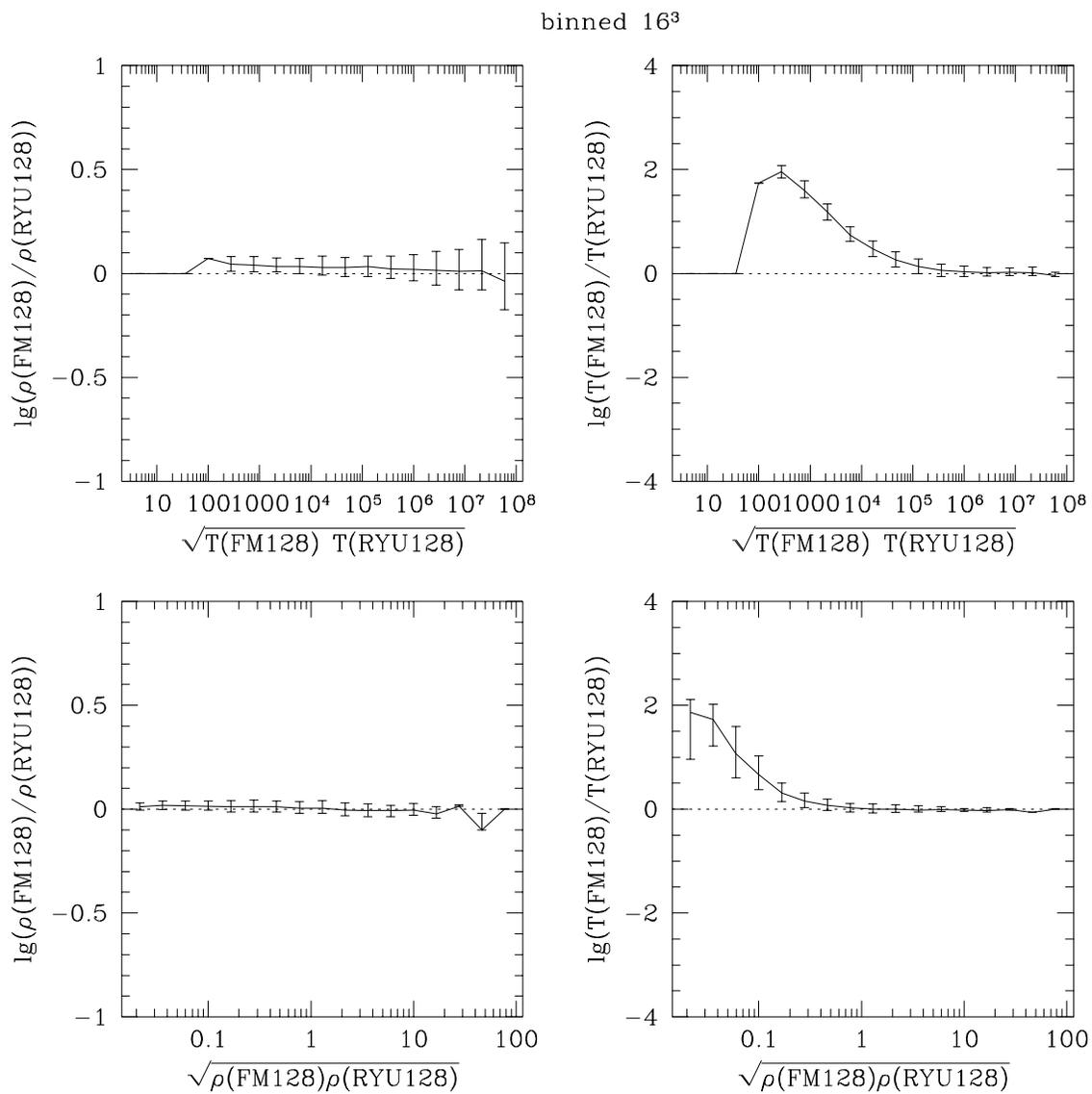}
\caption{Cosmological comparison for the $128^3$ fixed mesh.  The
error bars are the top and bottom quartiles of each bin.}
\label{fig:compfm}
\end{figure}

\begin{figure}
\plotone{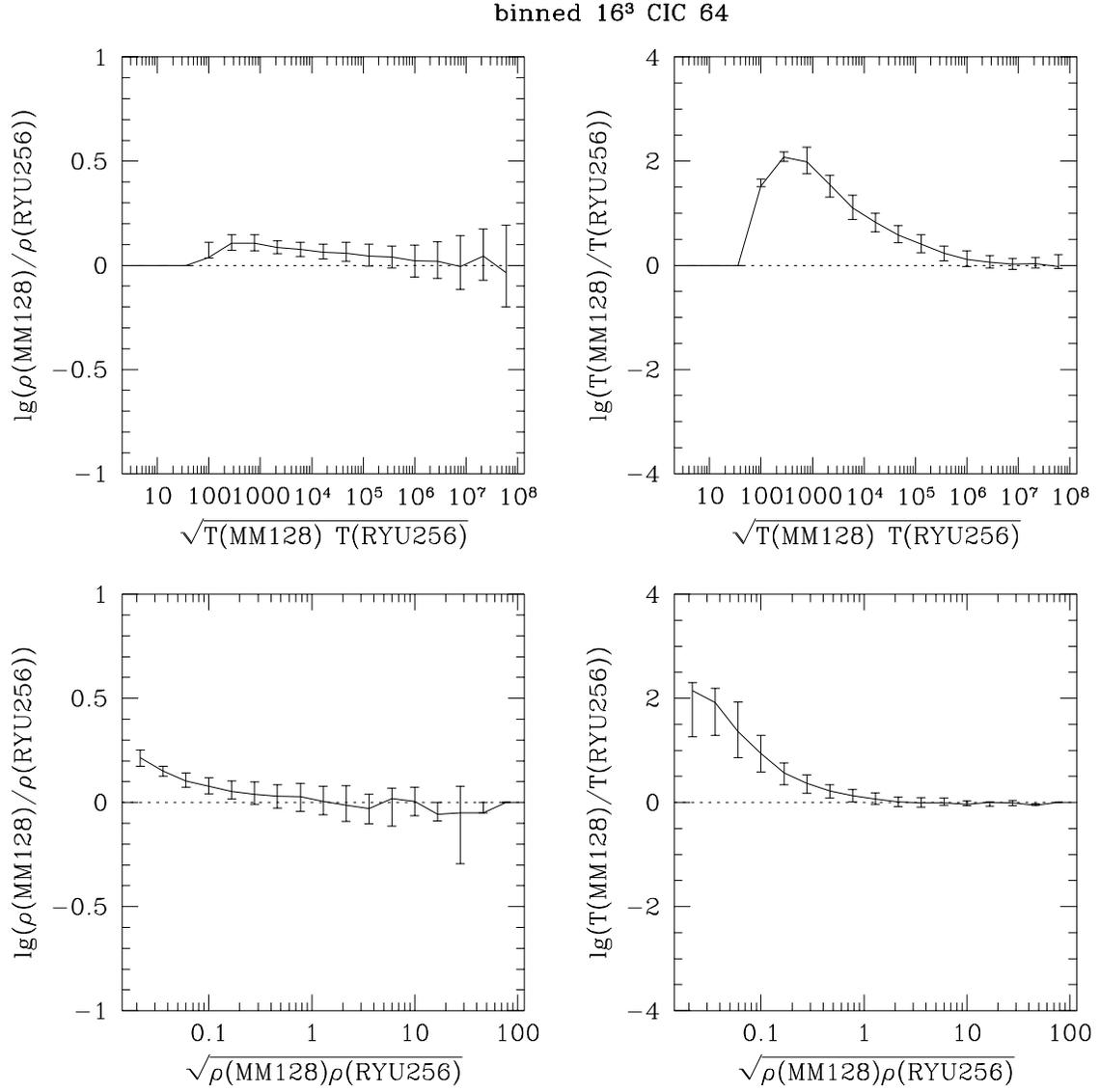}
\caption{Cosmological comparison for the $128^3$ moving mesh}
\label{fig:compmm}
\end{figure}

We also compare the results from out code to the standard test suite
in Kang et al (1994).  The density field for a $64h^{-1}$ Mpc box with
a $\sigma_8=1$ CDM normalization is given on an initial $64^3$ grid
linearly interpolated to $128^3$.  It is then evolved to the present
$(z=0)$ using a suite of different codes.  We compare convergence of
the final result binned into $16^3$ cells.  For the fixed mesh case,
we see in Figure \ref{fig:compfm} that the result agrees very well
with the Harten TVD scheme implemented by Ryu et al (1993b).
Agreement in temperature as a function of density (the lower right
graph) is poor in the low density regions, which are all unshocked and
thus should have very low temperature.  Since the current TVD scheme
does not incorporate any entropy variables, entropy is not conserved,
and a gas at very low temperature may experience sporadic heating and
cooling.  As we compare the moving mesh to the Kang \etal\ (1994)
simulations in Figure \ref{fig:compmm}, we obtain the expected result:
The agreement degrades in the low density regions where the grid
expands and the resolution degrades.  The increased resolution in the
high density regions is not visible when rebinned to such a course
grid.  We  loose some additional resolution because the
comparison is performed by mapping the moving mesh cells as constant
density Cloud-in-Cell particles on a fixed $64^3$ grid, which
introduces additional smoothing in the density field.

\begin{figure}
\plotone{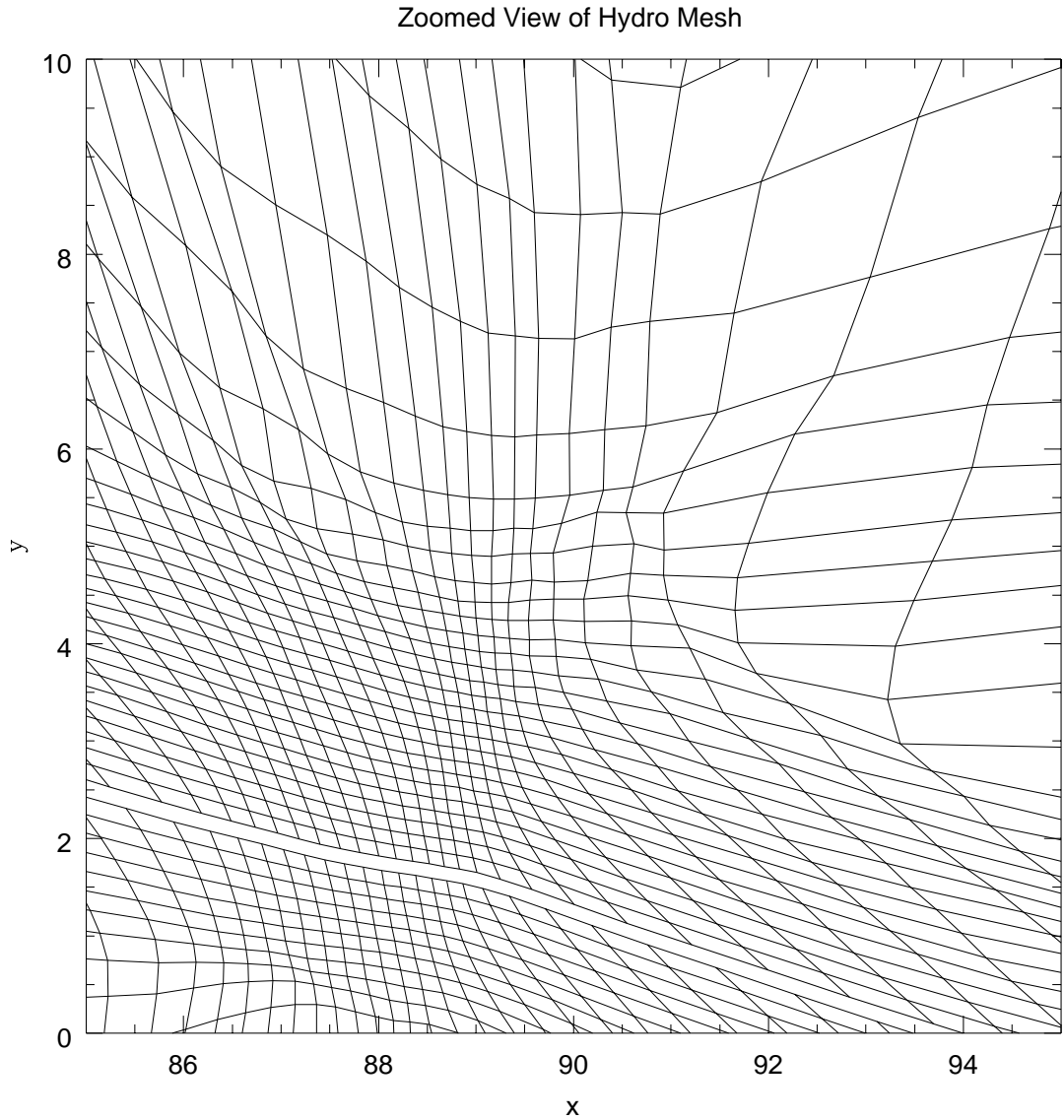}
\caption{A magnified view of a portion of Figure
\protect\ref{fig:mesh128}.
All grid lines are plotted.  The highest density regions are compression
limited at $\xi_m=0.1$.  In this state, the absence of rotation in the
coordinate system is apparent.  The empty channel running across the
graph is the grid periodicity boundary.  The lighter lines on the
lower part are the periodic image of the top region of the mesh.}
\label{fig:mesh128zoom}
\end{figure}

A slice of the $128^3$ mesh is shown in Figure \ref{fig:mesh128}.
Only every other grid line is represented in the graph.  The pancakes
are well represented, and the grid regularity is apparent.  In the
magnified view in Figure \ref{fig:mesh128zoom} the highest density
region is compression limited at $\xi_m=0.1$.  The grid reverts to a
regular Cartesian frame with normal orientation in these regions.
Since the grid equations are Galilean invariant, the high compression
region is allowed to move with a bulk motion to follow the fluid.

\section{Future Work}

Possible algorithmic improvement for the future includes the following:

1. implementing isolated (non-periodic) boundary conditions, such that
the code could be used for non-cosmological applications.

2. incorporating truly three dimensional flux limiters, especially the
Local Extrema Diminishing (LED) scheme, which would reduce the mass
diffusion problem.

3. it might be possible to implement a rigorously mass conserving
coordinate system, where the net mass flux is explicitly set to zero.
In such a system no mass diffusion could possibly occur, and the
Layzer-Irvine energy would be explicitly conserved.  The flux limiter
would now have to be applied directly to the deformation potential.
It is not entirely clear that this can be performed using only local
operators.

4. implementing higher accuracy hydrodynamics solvers, including
Essentially Non-Oscillatory (ENO) and Piecewise Parabolic Mesh (PPM)
algorithms.

\section{Conclusions}

We have presented a simple hydrodynamical algorithm which combines the
advantages of grid based finite volume flux conservative schemes with
the dynamic range of SPH Monte-Carlo Lagrangian schemes.  The
essential ingredients are a coordinate grid which tracks the potential
flow of the fluid, and a fast multigrid gravity solver.  By tracking
the potential flow, the mass per volume element remains constant,
giving an astrophysically desirable resolution which is roughly
constant in mass coordinates.  Furthermore, by following potential
flow and smoothing the grid and using limiters, the grid geometry
stays regular.  The curvilinear transformation maintains nearest
neighbor relations even for typical cosmological density contrasts of
$10^4$ and in the presence of vorticity in the fluid itself.  The full
curvilinear Eulerian equations of motion are solved on the grid, such
that even on a non-optimal or incorrect grid second order accurate
computation of hydrodynamic quantities would be assured.  On each grid
volume, the averaged conserved quantities (density, momenta, energy)
are stored, and at each time step, the flux between these control
volumes is computed to second order accuracy using the relaxing TVD
algorithm.  The equations in explicit flux conservative form guarantee
compliance with the Rankine-Hugionot shock jump conditions.

Gravitational force terms lead to energy violation in the presence of
numerical diffusion.  We have provided a compensation scheme which
nearly conserves the total energy even in the presence of such
diffusion.  By running a simulation with and without this
compensation, we can obtain a good estimator of the local errors.

The code runs very efficiently in terms of both memory and floating
point operations.  The current code parallelizes on symmetric
multiprocessor shared memory and  vector machines.  In Appendix
\ref{app:perf} we give accurate estimates of computational effort in
terms of floating point operations.

We have performed a large test suite on the code, and have
demonstrated the advantages for many cosmological problems.  On a
fixed grid the algorithm performance approaches that of other
state-of-the-art hydrodynamic schemes, and that accuracy is retained
when the mesh deforms strongly.  The coding is relatively short and
the algorithms simple.

The code is freely available to anyone for non-profit use.  Please
contact the author for more details.

We are grateful to J.P. Ostriker for all his support and ideas as well
as N. Gnedin, G. Xu and D. Spergel for helpful discussions.  This work
was supported in part by the NSF HPCC initiative under grant
ASC93-18185.

\appendix
\section{Curvilinear Coordinates}
\label{app:curvi}
Here we review some of the curvilinear transformations used in the
paper.

\subsection{Curvilinear Conservation Laws}

In this section we derive Equation (\ref{eqn:moving}).  We wish to
apply a general coordinate tranformation to an equation of the form
\begin{equation}
\left. \frac{\partial u}{\partial t}\right|_x + \frac{\partial
F^i[u]}{\partial x^i} = 0
\end{equation}
where $u=u(t,x)$ and the partial derivative with time holds $x$
fixed.  We express the time dependent coordinate transformation as
$x=x(\xi,t)$.  Applying the chain rule, we obtain
\begin{equation}
\left. \frac{\partial }{\partial t} \rg u\right|_\xi
+ \frac{\partial}{\partial \xi^\alpha}\left[ \rg e^\alpha_i
      ( F^i-u \dot{x}^i)\right]
- \left. \rg \frac{\partial }{\partial t}  u\right|_\xi
+ \rg u e^\alpha_i \dot{e}^i_\alpha - \left(F^i-u\dot{x}^i\right)
\frac{\partial }{\partial \xi^\alpha} (\rg e^\alpha_i)
= 0
\label{eqn:expand}
\end{equation}
where a dot indicates the partial derivative with respect to time
keeping $\xi$ fixed.
Expanding the determinant by Kramer's rule, we recall that $\partial_t\rg
= \rg e^\alpha_i\partial_t {e^i_\alpha} $, which eliminates the third and
fourth term in Equation (\ref{eqn:expand}).  We note that the triad
$e^\alpha_i$ is a one index contravariant vector.  Some algebra shows
that the quantity
\begin{equation}
\frac{1}{\rg} \partial_\alpha (\rg e^\alpha_i)
\label{eqn:scalar}
\end{equation}
is a scalar under coordinate transformations.  In Euclidian space,
expression (\ref{eqn:scalar}) vanishes everywhere.  Thus this scalar
is zero in all coordinates, and the last term in Equation
(\ref{eqn:expand}) is also identically zero.  QED.

\subsection{Eigenvalues}
We recall that the triad $e^i_\alpha$ is symmetric and positive
definite.  In the course of the computation, its eigenvalues will be
needed, which are computed as follows (CRC 1991):

1. let $A_{ij}=e^k_\alpha \delta^{\alpha i}\delta_{kj}$

2. let $t=$Trace$(A_{ij})$, \ \ $B_{ij}=A_{ij}-t/3$

3. let $a=B_{11}^2+B_{12}^2+B_{13}^2+B_{11}B_{22}+B_{22}^2+B_{23}^2$,
$m=2\sqrt{a/3}$

  $b= - B_{11}B_{12}^2 + B_{11}^2B_{22} - B_{12}^2B_{22} + B_{13}^2B_{22} + 
        B_{11}B_{22}^2 - 2B_{12}B_{13}B_{23} + B_{11}B_{23}^2$,

4. let $\theta=\cos^{-1}(3b/am)/3, \ \ r_1=\cos(\theta), \
r_2=\cos(\theta+2\pi/3), \ r_3=\cos(\theta+4\pi/3)$.

5. the eigenvalues are given by $\lambda_i=r_i+t/3$, which are given
in decreasing order.

The eigenvalues are of course real and positive.  The eigenvalues of
the inverse triad are the reciprocals of the eigenvalues of the triad.

Another set of eigenvalues which are needed to implement the relaxing
TVD algorithm are the eigenvalues of the Jacobian in equation
(\ref{eqn:moving}) in curvilinear coordinates.  As in Yee (1989) the
magnitude of the largest eigenvalue is given as max($\lambda_\alpha$),
where 
\begin{equation}
\lambda_\alpha = c \sqrt{\sum_{i=1}^3(e^\alpha_i)^2}
+\left|e^\alpha_i(v^i-\dot{x}^i)\right|
\end{equation}
and $c^2=\partial P/\partial\rho$ is the sound speed.

\subsection{Spherical Symmetry}

Assuming spherical symmetry n $N$ dimensions, some exact relations
between the metric and the 
deformation potential exist.  We define a parameter $\lambda = r^2$,
and let a prime denote differentiation with respect to $\lambda$.
The volume element is given as
\begin{equation}
\rg = \left(1+2\phi'\right)^{N-1}\left(1+2\phi'+4\lambda \phi''\right)
\label{eqn:sphrg}
\end{equation}
for an $N$-dimensional space.  We see in particular that in one
dimension, i.e. planar symmetry of three dimensions,
(\ref{eqn:sphrg}) becomes a linear 
equation.  For the next formulae we will assume that $N=3$.

If we prescribe the volume element, which corresponds to the grid
density, we wish to solve for the deformation potential.  It is given
by the following formula:
\begin{equation}
\phi = \int_0^r\left[ \ ^3\sqrt{3 \int_0^u v^2\rg dv}-u \right] du.
\label{eqn:sph3}
\end{equation}
In the case that $\rg$ is constant, (\ref{eqn:sph3}) simplifies to
\begin{equation}
\phi=\frac{r^2}{2}(\rg^{1/3}-1).
\end{equation}
We see that the deformation potential is  in general an inverted
parabola around density minima and a parabola around
density peaks.

For small fluctuations in density, let us enforce $\rho\rg=1$, and set
$\rho \equiv 1+\delta \rho, \ \delta \rg \equiv -\delta \rho$.  We
wish to approximate $\rg=|1+\phi_{,\alpha\beta}|\approx
1+\Nabla^2\phi+O(\phi^2)$.  For small fluctuations, we obtain
$\Nabla^2\phi=-\delta\rho$, and the deformation potential $\phi=-4\pi
GV$ is proportional to the gravitational potential.  We have recovered
the Zeldovich approximation with our displacement {\it ansatz} in
linear theory.

While the decomposition into potential flow and displacement is unique
for small density fluctuations (apart from boundary conditions), the
{\it ansatz} (\ref{eqn:ansatz}) is a particular choice in strongly
curvilinear coordinates.  It generalizes the notion of potential flow
by defining a frame through the symmetric triad in which no net
rotation of the coordinate system occurs.  The change is described by
a displacement and stretching of coordinate space alone.  We call this
rotation free moving frame the generalized potential flow for strongly
compressed gases.

\section{Relaxing TVD}
\label{app:relaxing}

The method is most easily illustrated in 1+1 dimensions.  Consider a
conservation equation of the form
\begin{equation}
\dot{u} + \partial_x F[u]=0
\label{eqn:sample}
\end{equation}
We replace that equation by another system
\begin{eqnarray}
\dot{u} +  \partial_x cv &=& 0
\nonumber \\
\dot{v} +  \partial_x cu &=& -\frac{1}{\epsilon} \left( v-F[u]
\right).
\label{eqn:relaxing}
\end{eqnarray}
where $c(x,t)$ is a free parameter called the {\it freezing speed}.
(\ref{eqn:relaxing}) is a linear advection equation with a non-linear
stiff source term.  The essence is to apply Strang splitting on these
two pieces.  A TVD flux/slope limiter is applied to the linear
advection equation, while an implicit backward Euler step enforces the
source term.  Xin and Jin (1994) showed this algorithm to be TVD
under the constraint that $c$ be greater than the characteristic speed
$\partial F/\partial u$.  One can now take the limit as $\epsilon
\longrightarrow 0$, which results in a {\it relaxed} algorithm.  Time
integration is implemented using a second order Runge-Kutta method.

To solve the linear part of (\ref{eqn:relaxing}), we decouple the
equations through a change of variables $w_1=u+v$ and $w_2=u-v$.  The
linear equation
\begin{equation}
\dot{w}+\partial_x cw=0
\end{equation}
is discretized in space using a Monotone Upstream-centered Scheme for
Conservation Laws (MUSCL) scheme.

We consider the conserved averaged quantities $w$ to be defined at
integer grid cells $x^n$.  Then we need to define the fluxes at cell
boundaries, ${\cal F}\equiv cw$ at $x^{n+1/2}$.
We then have $\partial_x cw = {\cal F}(x^{n+1/2})-{\cal F}(x^{n-1/2})$.  The
remaining trick is to define the flux $\cal F$ at half cells.

The first order upwind definition is simply ${\cal F}(x^{n+1/2}) =
cw(x^n)$, assuming flow is to the right.  There are two second order
choices: 1. 
$(cw(x^n)+cw(x^{n+1}))/2$ and 2. $(3/2)cw(x^n)-(1/2)cw(x^{n-1})$.  We
generalize the choices as ${\cal F}(x^{n+1/2}) = cw(x^n)+\Delta w$ where 
\begin{eqnarray}
\Delta w_+&=&\frac{cw(x^{n+1})-cw(x^n)}{2} \nonumber \\
\Delta w_-&=&\frac{cw(x^n)-cw(x^{n-1})}{2}.
\end{eqnarray}

Define the limiter minmod$(a,b)=($Sign$(a)+$Sign$(b))
$Min$(|a|,|b|)/2$.  It chooses the argument with smaller absolute
magnitude if the magnitudes have the same sign, and returns zero
otherwise.  The choice $\Delta w=$minmod$(\Delta w_+, \Delta w_-)$ is
the simplest of TVD MUSCL choices, which we use in our code.  Near
extrema of the flux vector the second order scheme reverts to a first
order upwind scheme.

The geometric interpretation is quite simple.  We start with the first
order upwind flux, and correct it using either left or right values,
choosing the one which demands a smaller correction.  If we are at an
extremum, the two corrections have opposite sign, and we do not
correct at all.  This approach is called a flux limiter.  For a
description of the mathematical justification, see for example Yee
(1988).  We have also implemented the whole range of TVD limiters
according to Hirsch (1990), of which the so-called ``superbee'' is the
least diffusive.

In the limit of a relaxed scheme where $\epsilon=0$, we operate with
the constraint that $v=F[u]$ at the beginning of each partial step,
and $v$ becomes only an auxiliary vector to calculate flux limiters.
The advantage of the relaxed scheme is that it requires no knowledge
of the eigenvectors or eigenvalues of the flux function, only an
estimate for the lower bound of the maximum eigenvalue.  Since the
curvilinear equations of motion (\ref{eqn:moving}) are rather complex,
this is of computational advantage.  It is also quite simple to
implement.  The time step is limited by the freezing speed $c$, and we
obtain simple expressions to compute the correct time step.

In several spatial dimensions, the simplest generalization is to apply
the freezing advection to each dimension.  To illustrate in two
dimensions, we start with
\begin{equation}
\dot{u} + \partial_x F + \partial_y G = 0
\end{equation}
which we convert into the relaxing equation
\begin{eqnarray}
\dot{u} +  \partial_x cv +  \partial_y cw&=& 0
\nonumber \\
\dot{v} +  \partial_x cu &=& -\frac{1}{\epsilon} \left( v-F[u] \right)
\nonumber \\
\dot{w} +  \partial_y cu &=& -\frac{1}{\epsilon} \left( w-G[u] \right).
\label{eqn:set}
\end{eqnarray}

The limiter is then applied to each pair $(u,v)$ and $(u,w)$.  By
applying the Runge-Kutta time integrator to the whole system, the
algorithm is not dimensionally split.  One should note, however, that
the slope limiter is in fact dimensionally split.  This could be
circumvented by using a Local Extrema Diminishing (LED)
limiter on the whole set of linear advection equations (\ref{eqn:set}).
We have not implemented this method, since it has a large operation
count and program complexity.  In the current numerical experiments, no
direct problems with the directional slope limiter has been observed.

While not as rigorous or accurate as the Harten or PPM scheme when
applied to the full three dimensional system, the relaxed scheme
offers simplicity and robustness.  For a description of curvilinear
TVD schemes, see Yee (1987).

\section{Performance Issues}
\label{app:perf}
\subsection{Memory}

The code is extremely memory friendly.  The minimal required storage
count for arrays of size $N^3$ is 7: 5 for the hydrodynamic arrays, 1
for the deformation potential $\phi$, 1 for the time derivative of the
deformation potential $\dot{\phi}$ and 1 for the gravitational
potential $V$.  In principle the last two could be stored in the same
array since they are not in principle needed simultaneously.  Some
additional storage of order $N^3/2$ would be required for the
multigrid scheme.

In practice, we use 10 arrays in the current implementation.  The
gravitational potential is stored twice, allowing us to linearly
interpolate from the previous two time steps as an initial guess to
the multigrid gravity solver.  An extra array is used to store $\delta
\rho\rg=(\rho-1)\rg$.

The program allocates up to 5 more arrays for special purposes.  An
additional array is needed for computing N-body particles to second
order accuracy, which is used to store the deformation potential of a
previous time step.  Four more arrays are needed to implement the
gravitational energy conservation scheme.

All the temporary arrays that are needed for the Runge-Kutta scheme
and the relaxation to proceed efficiently without recomputation of old
values are all stored in two dimensional arrays.

\subsection{Parallelization}

All operations except for the gravity solver are explicitly performed
on a regular grid, and would thus parallelize straightforwardly on any
kind of parallel or vector machine with no load imbalance issues.  The
parallel multigrid algorithm has been investigated in detail in the
literature, and can in principle be performed efficiently on a
parallel machine.

The current code runs in parallel on a shared memory SGI Power
Challenge with 150 MHz R8000 processors.  In order to maintain
simplicity and storage efficiency for the temporary two dimensional
arrays, most parallelization is done at the second level of loops.  On
a machine with 8 Processors, we obtain a speedup of 6.

\subsection{Operation Count}

The most direct objective measure of computation speed is the floating
point count.  For the moving mesh, this count is independent of
clustering or deformation, and thus an accurate predictor of the
execution time.  The actual sustained floating point speed of each
machine depends on many parameters, including compiler version and
many compiler fine tuning options.

Let us define the basic operating cost $C$ as the number of floating
operations per cell per Courant step.  This is an objective measure of
any numerical code, and depends only on the physical parameters and
grid resolution.  The code currently takes two hydrodynamic and one
gravity time step in each Courant time interval.  It also calls the
multigrid solver four times, twice for the deformation potential and
twice for the gravitational potential.  The gravitational solver is
called twice in a row to provide a minimal gravitational error.
Errors in the deformation potential do not enter directly in any other
error estimates, and it is thus not crucial to solve the deformation
potential accurately.

We thus have $C=\sqrt{3}(2C_H+4C_M)$, where $C_H$ is the cost per cell
per time step for the curvilinear relaxing TVD hydro, and $C_M$ is the
cost per cell of the multigrid solver.  In Pen (1995) we computed the
asymptotic cost of $C^a_M=918$ for a grid of infinite size.  The
actual count obtained for a $64^3$ grid using the floating point
counter on a Cray C90 is $C_M=1419$.  The discrepancy arises in part
because the estimate of $C^a_M$ was based on computing a metric
tensor, and the sum of relaxation sweeps over all subgrid.  To test
the compiler count on these operations alone, we obtain a value of
$1285$ from the hardware counter.  While there are certain neglected
costs, in particular that of edge effects and other costs which are of
$O(N^2)$, the discrepancy is much larger than they could account for.
We have no explanation for the difference.  When the relaxation
routine alone is timed, the hardware counter obtains 52 additions, 33
multiplications and one division, while a direct source statement
count yields 41 additions, 26 multiplications and 1 division.  On the
SGI compiler version 6.0.2, we have analyzed the generated assembly
code, which obtained a 15\% higher floating point count than the
original source code.  This was possibly due to algebraic
rearrangements which may have improved instruction scheduling.  In
addition, the cost of prolongation and projection operators, as well
as operations which force a volume weighted zero sum was neglected.
These may account for the remaining $134$ operations.

For the hydrodynamic calculations, we only have the hardware count
data available.  Using a $64^3$ run, we obtain $C_H\approx 2622$.
This yields a total operation count $C\approx 19000$, of which half
arises from the hydrodynamics, and half from the multigrid solver.  On
a single CPU of a Cray C90 the code achieves about 300 Mflop on
$256^3$ runs out of a theoretical peak of 1Gflop, which is 30\% of
peak.  The vector lengths for the multigrid relaxation are half the
box widths, which means we fill the vector length using a $256^3$
mesh.  On an SGI power challenge R8000, the code currently achieves 48
Mflop out of a theoretical peak speed of 300 Mflop on a $64^3$ run.
Cache misses account for about 23\% of the computing time on the SGI.

\clearpage

\begin{figure}
\plotone{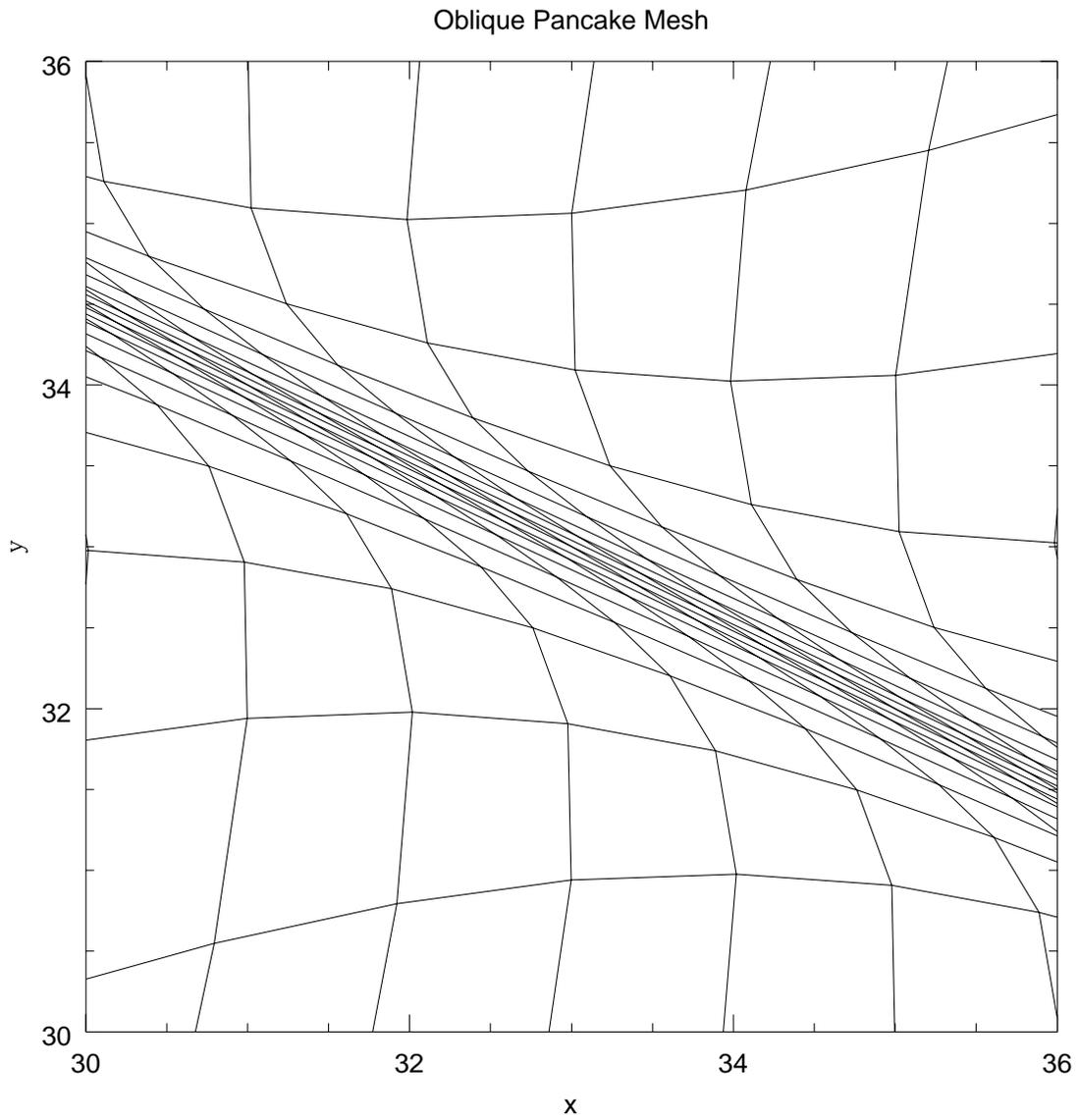}
\caption{The mesh on which Figure \protect\ref{fig:oblique} was
computed.  We see that in the densest regions, the grid is highly
oblique.  The numerical grid neighbors are no longer the nearest
physical neighbors.  Despite such extreme grid distortions, the
solution remains well behaved.
}
\label{fig:omesh}
\end{figure}

\end{document}